\documentclass[aps,pra,twocolumn,nopacs,floatfix]{revtex4-2}
\usepackage{epsfig}
\usepackage{graphicx}
\usepackage{dcolumn}
\usepackage{amsthm,amsmath,amsfonts}
\usepackage{mathtools}
\usepackage{comment}
\usepackage[colorlinks]{hyperref}
\usepackage{xcolor}
\usepackage{orcidlink}
\usepackage{rotating}
\usepackage[normalem]{ulem}

\begin{document}

\title{Investigating Roles of Triple Excitations for High-precision Determination of Clock Properties of Alkaline Earth Metal Singly Charged Ions}

\author{A. Chakraborty$^{a}$ \orcidlink{0000-0001-6255-4584}}
\email{arupc794@gmail.com}

\author{Vaibhav Katyal$^{a,b}$ \orcidlink{0000-0002-7717-8558}}

\author{B. K. Sahoo$^{a}$ \orcidlink{0000-0003-4397-7965}}
\email{bijaya@prl.res.in}
\affiliation{$^a$Atomic, Molecular and Optical Physics Division, Physical Research Laboratory, Navrangpura, Ahmedabad 380009, India\\
$^b$Indian Institute of Technology Gandhinagar, Palaj, Gandhinagar 382355, India}

\begin{abstract}
High-accuracy calculations of electric dipole polarizabilities and quadrupole moments ($\Theta$) of the clock states of the singly charged calcium (Ca$^+$), strontium (Sr$^+$) and barium (Ba$^+$) alkaline-earth ions are estimated by employing relativistic coupled-cluster (RCC) theory. It demonstrates importance of the triple excitations in the RCC method for precise determination of the above quantities. We also observe a different trend of correlations in the $\Theta$ values than an earlier study with respect to orbitals from higher angular momenta. Reliability of the results is verified by comparing the calculated energies, magnetic dipole hyperfine structure constants, and lifetimes of the atomic states with the experimental values of the $^{43}$Ca$^+$, $^{87}$Sr$^+$ and $^{137}$Ba$^+$ ions. Nuclear quadrupole moments of these isotopes are also estimated by combining calculations with the measured electric quadrupole hyperfine structure constants, showing large deviations from the literature values.   
\end{abstract}

\date{\today}

\maketitle 

Singly charged alkaline-earth metal ions are instrumental in investigating many fundamental physics. Transitions between the ground ($ns$) and metastable $(n-1)d ~ ^2D_{3/2,5/2}$ states, with principal quantum number $n$, in the heavier alkaline ions play key roles in the clock frequency measurements \cite{Olmschenk2009, Sherman2008, Sahoo2009-3}, probing atomic parity violation (APV) \cite{Fortson1993, Wansbeek2008, Sahoo2007} and analyzing isotope shift (IS) data to extract nuclear charge radii \cite{Sahoo2009, Gossel2013, Ruiz2016} and mass of a possible new vector boson from the non-linear effects in the King's plot \cite{Berengut2018}, constructing qubits \cite{Dutta2022, Clark2021, Moller2007} etc. to name a few. Owing to laser-accessible energy level structures, many of these ions can be efficiently manipulated with lasers, allowing for long interrogation times and thereby enabling high-precision measurements. In fact, a recent proposal aims to measure the APV amplitude in Ba$^+$ to 0.1\% precision using an ion trap technique \cite{Craik2025}.

From the perspective of carrying out above studies, high-accuracy calculations score as much as the precision measurements. It, therefore, necessitates to underscore roles of higher-order electron correlation effects in different atomic properties for more reliable outcomes. The prominent systematic effects in the high-precision atomic experiments are Stark shifts, blackbody radiation shifts, quadrupole shifts, and Zeeman shifts. In this regard, independent experiments are required to set-up to measure these systematic effects to the intended precision. Sometimes managing these experiments could be extremely challenging, time consuming and expensive. In such situation, high-accuracy atomic calculations to estimate systematics can be the alternative. In view of this, precise knowledge of electric dipole polarizabilities ($\alpha_d$) and quadrupole moments ($\Theta$) of the clock states of the Ca$^+$, Sr$^+$ and Ba$^+$ ions are of immense interest.

\begin{table*}[t!]
\setlength{\tabcolsep}{3pt}
\caption{Calculated energies (in cm$^{-1}$) and $\alpha_d^S$ values (in a.u.) of the clock states of Ca$^+$, Sr$^+$, and Ba$^+$ at the DHF and RCC methods. The final values are compared with experimental values (Expt.) and other precise calculations.}
\centering
\begin{tabular}{l ccc ccc ccc}
\hline\hline
 & \multicolumn{3}{c}{Ca$^+$} & \multicolumn{3}{c}{Sr$^+$}  & \multicolumn{3}{c}{Ba$^+$}\\
\cline{2-4}\cline{5-7} \cline{8-10} \\
Method & $4s ~ ^2S_{1/2}$ & $3d ~ ^2D_{3/2}$ & $3d ~ ^2D_{5/2}$ & $5s ~ ^2S_{1/2}$ & $4d ~ ^2D_{3/2}$ & $4d ~ ^2D_{5/2}$ & $6s ~ ^2S_{1/2}$ & $5d ~ ^2D_{3/2}$ & $5d ~ ^2D_{5/2}$ \\
\hline 
\multicolumn{10}{c}{Energies in cm$^{-1}$} \\
DHF    & 91440.02 & 72618.65 & 72594.55 &  84041.96 & 67384.80 & 67241.97 & 75339.35 & 68139.75 & 67665.52 \\
RCCSD  & 95906.31 & 82293.80 & 82212.33 & 89094.93  & 74404.68 & 74101.58 & 80883.80 & 75738.94 & 74907.46 \\
RCCSDT & 95681.86 & 81912.20 & 81836.51 & 88859.83 & 74239.34 & 73953.86 & 80482.23 & 75420.56 & 74629.95  \\
$+$Basis & 10.34 & 77.39 & 77.14 & 16.57 & 81.17 & 79.88  & 30.06 & 125.91 & 121.71 \\
$+$Breit & $-7.33$ & 37.67 & 52.85 & $-10.25$ & 41.74 & 58.52 & $-10.76$ & 51.55 & 74.77 \\
$+$VP  &  0.92 & $-0.38$ & $-0.38$  &  3.32 & $-0.90$ & $-0.88$ & 7.02 & $-2.19$ & $-2.07$ \\[1ex]
Final &  95685.79 & 82026.88 & 81999.12 & 88869.47 & 74391.48 & 74091.38 & 80508.55 & 75595.83 & 74824.36 \\[0.5ex]
Expt. \cite{NIST} &  95751.87 & 82101.68 & 82040.99 & 88965.18 & 74409.28 & 74128.94 & 80686.30 & 75812.448 & 75011.493 \\
$\Delta$(\%) & 0.07 & 0.09 & 0.05 & 0.11 & 0.02 & 0.05 & 0.22 & 0.29 & 0.25 \\
\hline 
\multicolumn{10}{c}{$\alpha_d^S$ values in a.u.} \\
DHF    & 98.90 & 104.92 & 102.24 & 127.03 & 148.90 & 139.42 & 185.52 & 91.29 & 88.78 \\
RCCSD  & 76.33  & 32.01 & 31.87 & 92.05 & 64.53 & 63.32 & 125.37 & 52.45 & 52.56  \\
$+$Basis & $-0.02$ & $-0.14$ & $-0.14$ & $-0.05$ & $-0.28$ & $-0.27$ & $-0.12$ & $-0.19$ & $-0.18$ \\
$+$Breit & $\sim 0.0$ & $-0.18$ & $-0.22$ & $-0.01$ & $-0.52$ & $-0.57$ & $\sim 0.0$ & $-0.23$ & $-0.28$ \\
$+$VP  & $-0.03$ & $-0.03$ & $-0.03$ & $-0.03$ & $-0.05$ & $-0.04$ & $-0.10$ & $-0.04$ & $-0.04$ \\[1ex]
Final & 76.3(5) & 31.7(5) & 31.5(5) & 92.0(7) & 63.7(8) & 62.4(8) & 125.2(9) &  52.0(8) & 52.1(8) \\[0.5ex]
Theory & 76.1(5)$^a$ & 32.0(3)$^a$ & 31.8(3)$^a$ & 88.29(1.0)$^c$ & 61.43(52)$^c$ & 62.87(75)$^c$ & 124.26(1.0)$^c$ & 48.81(46)$^c$ & 50.67(58)$^c$ \\
 & 73.0(1.5)$^b$ & 28.5(1.0)$^b$ & 29.5(10)$^b$ & 91.3(9)$^d$ &  &  62.0(5)$^d$ & 124.15$^e$ & &  \\
Expt. & & & & & & & 123.88(5) \cite{Snow2007} & &  \\
           & & & & & & & 125.5(10) \cite{Gallagher1982} & &  \\
\hline\hline\\
\end{tabular}
\\ $^a$\cite{Safronova2011}; $^b$\cite{Sahoo2009-2}; $^c$\cite{Sahoo2009-3}; $^d$\cite{Jiang2009} ;$^e$\cite{Tchoukova2008}
\label{tab1}
\end{table*}

Over the past decades, several experimental techniques like resonant excitation, Stark ionization and single ion trapping methods have been applied to measure transition probabilities, lifetimes, and hyperfine structure constants of the clock states of the Ca$^+$, Sr$^+$ and Ba$^+$ ions \cite{Barton2000, Kreuter2005, Mannervik1999, Biemont2000, Letchumanan2005, Yu1997, Gurell2007, Madej1990, Madej1990-2, Arbes1994, Silverans1991, Kurth1995, Nortershauer1998, Benhelm2007, Buchinger1990, Sunaoshi1993, Barwood2003, Becker1968, Silverans1986}. However, only a few limited measurements of $\alpha_d$ and $\Theta$ of the considered ions are available in the literature \cite{Snow2007, Gallagher1982, Roos2006, Barwood2004, Shaniv2016}. Again, poor agreement between these values with the calculations reported using different theories are being observed \cite{Roos2006, Barwood2004, Shaniv2016, Jiang2008, Itano2006, Sur2006}. This obscures improving accuracy of the clock transitions of the aforementioned ions. The {\it ab initio} calculations of $\alpha_d$ were performed using lower-order methods, while other values were estimated in the semi-empirical approaches by adopting a mixture of methods utilizing some of the dominantly contributing electric dipole (E1) matrix elements from the calculations along with the experimental energies \cite{Sahoo2009-3, Safronova2011, Tchoukova2008, Jiang2009, Sahoo2009-2}. On the other hand, the $\Theta$ values were estimated using the relativistic coupled-cluster (RCC) method \cite{Jiang2008, Sur2006} and multi-configuration Dirac Hartree-Fock (MCDHF) method \cite{Itano2006} approximating at the singles and doubles excitation levels. It has also been noticed that the extracted nuclear quadrupole moments ($Q_n$) of $^{43}$Ca, $^{87}$Sr and $^{137}$Ba from the hyperfine structure studies of these atoms or their ions substantially differ from each other \cite{Sahoo2009CaSrBa, Tangarxiv, Yu2004, Pyykko2001}, which needs to be addressed. 
 
\begin{sidewaystable}[htbp]
\setlength{\tabcolsep}{1pt}
\caption{$\alpha_d^{T}$ and $\Theta$ values (both are in a.u.) of the metastable states of Ca$^{+}$, Sr$^{+}$, and Ba$^{+}$ from different contributions of the present calculation and their comparisons with the available experimental values (Expt.) and accurate calculations.}
\centering
\begin{tabular}{l cc cc cc cc cc cc}
\hline\hline
 & \multicolumn{4}{c}{Ca$^+$} & \multicolumn{4}{c}{Sr$^+$}  & \multicolumn{4}{c}{Ba$^+$}\\
\cline{2-5}\cline{6-9} \cline{10-13} \\
Method & \multicolumn{2}{c}{$3D_{3/2}$} & \multicolumn{2}{c}{$3D_{5/2}$} & \multicolumn{2}{c}{$4D_{3/2}$} & \multicolumn{2}{c}{$4D_{5/2}$} & \multicolumn{2}{c}{$5D_{3/2}$} & \multicolumn{2}{c}{$5D_{5/2}$} \\
    & $\alpha_d^T$ & $\Theta$ & $\alpha_d^T$ & $\Theta$ &  $\alpha_d^T$ & $\Theta$ & $\alpha_d^T$ & $\Theta$ & $\alpha_d^T$ $\Theta$ & $\alpha_d^T$ & $\Theta$ \\
\hline \\
DHF & $-66.39$ & 1.712 & $-90.68$ & 2.451 &  $-93.83$ & 2.469 &  $-118.32$ & 3.559 & $-45.50$ & 2.733 & $-58.63$ & 3.994 \\ 
RCCSD & $-17.05$ & 1.288 & $-23.99$ & 1.846 & $-35.72$ & 2.038 & $-48.02$ & 2.948 & $-22.45$ & 2.278 & $-30.14$ & 3.353 \\
RCCSDT & - & 1.296 & - & 1.857 & - & 2.037 & - & 2.984 & - & 2.279 & - & 3.354 \\
$+$Basis & 0.10 & $-0.003$ & 0.14 & $-0.004$ & 0.19 & $-0.005$ & 0.26 & $-0.007$ & 0.12 & $-0.006$ & 0.16 & $-0.009$ \\
$+$Breit & 0.14 & $-0.002$ & 0.22 & $-0.003$ & 0.39 & $-0.002$ & 0.53 & $-0.006$ & 0.18 & $-0.003$ & 0.26 & $-0.006$ \\
$+$VP  & 0.03 & $\sim 0.0$ & 0.03 & $\sim 0.0$ & 0.04 & $\sim 0.0$ & 0.04 & $\sim 0.0$ & 0.04 & $\sim 0.0$ & 0.04 & 0.001 \\[1ex]
Final & $-16.8(2)$ & 1.291(6) & $-23.6(3)$ & 1.850(9) & $-35.1(4)$ & 2.030(8) & $-47.2(5)$ & 2.971(10) & $-22.1(2)$ & 2.270(9) & $-29.7(3)$ & 3.339(12) \\[0.5ex]
Theory & $-17.43(23)^{a}$ & 1.289(11)$^{c}$ & $-24.51(29)^{a}$ & 1.849(17)$^{c}$ & $-35.42(25)^{e}$ & 2.029(12)$^{c}$ &  $-48.83(30)^{e}$ & 2.935(17)$^{c}$ & $-24.62(28)^{e}$ & 2.256(11)$^{c}$ & $-30.85(31)^{e}$ & 3.319(15)$^{c}$  \\
& $-15.8(7)^{b}$ & 1.338$^{d}$ & $-22.45(5)^{b}$ & 1.917$^{d}$ &  & 2.107$^{d}$ &  $-47.7(3)^{f}$ & 3.048$^{d}$ &  & 2.297$^{d}$  &  & 3.379$^{d}$ \\
Expt. &  &  &  & 1.83(1) \cite{Roos2006} & & &  & 2.6(3) \cite{Barwood2004},~2.973$^{0.026}_{0.033}$\cite{Shaniv2016} &  & & \\
\hline\hline\\
\end{tabular}
$^a$\cite{Safronova2011};  $^b$\cite{Sahoo2009-2}; $^c$\cite{Jiang2008}; $^d$\cite{Itano2006}; $^e$\cite{Sahoo2009-3}; $^f$\cite{Jiang2009}
\label{tab2}
\end{sidewaystable}

In this work, we intend to perform first-principle calculations of $\alpha_d$ and $\Theta$ of the clock states of Ca$^{+}$, Sr$^{+}$, and Ba$^{+}$ by employing the recently developed singles, doubles and triples approximated RCC (RCCSDT) method. In order to validate the calculations, we present energies and magnetic dipole hyperfine structure constants ($A_{hf}$) of the clock states using the RCCSDT method and compare them with the experimental values. We also extract the $Q_n$ values of the $^{43}$Ca, $^{87}$Sr and $^{137}$Ba isotopes by combining the precisely measured electric quadrupole hyperfine structure constants ($B_{hf}$) with their corresponding calculations of the above ions. We consider as many as 40, 39, 38, 37, 36, 35, and 34 Gaussian-type orbital (GTO) basis functions for constructing Dirac-Hartree-Fock (DHF) orbitals of the $s$, $p$, $d$, $f$, $g$, $h$, and $i$ symmetries, respectively. The GTO parameters are optimized such that energies of core orbitals agree reasonably well with the numerical values from GRASP \cite{grasp}. Contributions from the orbitals belonging to orbital angular momentum $l\ge7$ (quoted as $+$Basis) are extracted using the third-order many-body perturbation (MP) theory (refer to Supplemental Materials \cite{suppl} for details). To account electron correlation effects, we have considered the Dirac-Coulomb (DC) Hamiltonian ($H_{DC}$) in the RCCSDT method, while contributions from the Breit interactions (given as $+$Breit) and lower-order vacuum polarization (VP) effects (given as $+$VP) are estimated at the singles and doubles approximation in the RCC (RCCSD) method \cite{Ginges2016, Sahoo2016} to improve accuracy of the calculations. Below, we present results from the DHF, RCCSD and RCCSDT methods to demonstrate how electron correlation effects arising through the triple excitations play important roles in improving the results. 

\begin{table}[t!]
\setlength{\tabcolsep}{3pt}
\caption{Demonstration of triples contributions to the $\alpha_d^S$ and $\alpha_d^T$ values (in a.u.) of the ground states of the three considered ions and metastable states of Ca$^+$.}
\centering
\begin{tabular}{l ccc cc}
\hline\hline
Quantity  & \multicolumn{3}{c}{Ca$^+$} & Sr$^+$ & Ba$^+$ \\
  \cline{2-4} \\
   & $4s ~ ^2S_{1/2}$ & $3d ~ ^2D_{3/2}$ & $3d ~ ^2D_{5/2}$ & $5s ~ ^2S_{1/2}$ & $6s ~ ^2S_{1/2}$ \\
\hline \\
$\alpha_d^S$   & $-0.25$ & $0.42$ & $0.40$  & $-0.47$ & $-0.08$ \\
$\alpha_d^T$   & $0.0$ & $-0.31$ & $-0.30$ & $0.0$ & $0.0$ \\
\hline\hline\\
\end{tabular}
\label{tab12}
\end{table}

\begin{table*}[t!]
\setlength{\tabcolsep}{0.5pt}
\caption{Reduced E2 and M1 amplitudes (in a.u.) of the clock transitions in Ca$^+$, Sr$^+$, and Ba$^+$ at different levels of calculations and their comparisons with previously available precise calculations. Numbers appearing as $a[b]$ mean $a\times10^b$.}
\centering
\begin{tabular}{l ccc ccc ccc}
\hline\hline
 & \multicolumn{3}{c}{Ca$^+$} & \multicolumn{3}{c}{Sr$^+$}  & \multicolumn{3}{c}{Ba$^+$}\\
\cline{2-4}\cline{5-7} \cline{8-10} \\
Method & $S_{1/2}-D_{3/2}$ & $S_{1/2} - D_{5/2}$  & $D_{3/2} - D_{5/2}$ & $S_{1/2}- D_{3/2}$ & $S_{1/2} - D_{5/2}$  & $D_{3/2} - D_{5/2}$ & $S_{1/2} - D_{3/2}$ & $S_{1/2} - D_{5/2}$  & $D_{3/2} - D_{5/2}$ \\
\hline 
\multicolumn{10}{c}{E2 amplitudes} \\
DHF    & 9.767 & 11.978 & 5.018 & 12.969 & 15.973 & 7.261 & 14.764 & 18.385 & 8.092 \\
RCCSD  & 7.914 &  9.716 & 3.777 & 11.142 & 13.762 & 6.005 & 12.662 & 15.850 & 6.777 \\
RCCSDT & 7.959 & 9.770 & 3.803 & 11.159  & 13.779 & 6.008 & 12.700 & 15.890 & 6.797 \\
$+$Basis & $-0.011$ & $-0.013$ & $-0.008$ & $-0.017$ & $-0.020$ & $-0.014$ & $-0.029$ & $-0.036$ & $-0.019$\\
$+$Breit & $-0.006$ & $-0.009$ & $-0.005$ & $-0.007$ & $-0.014$ & $-0.008$ & $-0.009$ & $-0.020$ & $-0.010$ \\
$+$VP  & $\sim 0.0$ & $\sim 0.0$ & $\sim 0.0$ & $\sim 0.0$ & $\sim 0.0$ & $\sim 0.0$ & $\sim 0.0$ & $\sim 0.0$ & $\sim 0.0$ \\[1ex]
Final &  7.942(7) & 9.748(11) & 3.790(5) & 11.135(15) & 13.745(20) & 5.986(7) & 12.662(20) & 15.834(25) & 6.768(9) \\[0.5ex]
Theory \cite{Kaur2021} & 7.78(4) & 9.56(4) & 3.68(2) & 11.01(4) & 13.60(5) & 5.90(2) & 12.49(11) & 15.65(14) & 6.65(6) \\
Theory \cite{Safronova2017} & 7.94(4) & 9.750(47) &  & 11.13(39) & 13.745(29) &  & 12.63(11) & 15.800(79) & \\
\hline 
\multicolumn{10}{c}{M1 amplitudes} \\
DHF    & $-4.1[-7]$ &  & 1.549 &  $1.6[-6]$ & & 1.549 & $5.5[-6]$ & & 1.549 \\
RCCSD  & $3.5[-4]$ &  & 1.555 & $3.0[-5]$ & & 1.553 & $3.1[-4]$ & & 1.554  \\
RCCSDT & $3.5[-4]$ &  & 1.555 & $3.2[-5]$ & & 1.553 &  $3.2[-4]$ &  & 1.554 \\
$+$Basis    & $\sim 0.0$ &  & $\sim 0.0$ & $\sim 0.0$ & & $\sim 0.0$ & $\sim 0.0$ & & $\sim 0.0$\\
$+$Breit & $-1.1[-5]$ &  & $\sim 0.0$ & $1.1[-5]$ &  &  $-0.002$ & $1.0[-5]$ & & $\sim 0.0$\\
$+$VP  & $\sim 0.0$ & & $\sim 0.0$ & $\sim 0.0$ &  & $\sim 0.0$ &  $\sim 0.0$ &  & $\sim 0.0$ \\[1ex]
Final &  $3.4(2)[-4]$  &  & 1.555(2)  & $4.3(3)[-5]$ & & 1.551(2) & $3.3(3)[-4]$ & & 1.554(2) \\[0.5ex]
Theory \cite{Kaur2021} & $\sim 0.0$ &  & 1.5491 & $\sim 0.0$ &  & 1.5492 & $0.0002$ &  &  1.5493 \\
\hline\hline\\
\end{tabular}
\label{tab3}
\end{table*}

\begin{table}[t!]
\setlength{\tabcolsep}{0.7pt}
\centering
\caption{Estimated lifetimes of clock states of the considered ions computed using the experimental wavelengths (shown with $*$), their comparisons with the literature values. Transition rates from the M1 and E2 modes are also given.}
\begin{tabular}{cc cc cc} \hline \hline
Atomic & Decay  &  Mode & $A_{n\rightarrow k}^o$ & \multicolumn{2}{c}{$\tau_n$ (in s)}  \\ 
\cline{5-6} \\
 State  & State &   & in s$^{-1}$  & Theory & Expt. \\ 
\hline 
\multicolumn{6}{l}{\underline{Ca$^+$ ion}} \\ 
$3D_{3/2}$ & $4S_{1/2}$ & M1 & $1.98[-6]$ & 1.195(4)$^*$ & 1.20(1) \cite{Barton2000} \\
               &                & E2 & 0.83690 & 1.194(11) \cite{Safronova2017} & 1.176(11) \cite{Kreuter2005} \\
               &                & &  & 1.24(1) \cite{Kaur2021} &  \\ [1ex]
$3D_{5/2}$ & $3D_{3/2}$ & M1 & $2.43[-6]$ & 1.164(5)$^*$ & 1.168(7) \cite{Barton2000}  \\
               &                & E2 & $2.21[-13]$ & 1.163(11) \cite{Safronova2017} & 1.168(9) \cite{Kreuter2005} \\
               & $4S_{1/2}$ & E2 & 0.85936 & 1.21(1) \cite{Kaur2021} & \\[0.3ex]
\multicolumn{6}{l}{\underline{Sr$^+$ ion}} \\ 
$4D_{3/2}$ & $5S_{1/2}$ & M1 & $3.84[-8]$ & 0.441(2)$^*$ & 0.435(4) \cite{Mannervik1999,Biemont2000} \\
               &                & E2 & 2.26822 & 0.437(14) \cite{Safronova2017} & 0.455(29) \cite{Biemont2000} \\
               &                & &  & 0.451(3) \cite{Kaur2021} &  \\ [1ex]
$4D_{5/2}$ & $4D_{3/2}$ & M1 & $0.24[-3]$ & 0.394(2)$^*$ & 0.408(22) \cite{Biemont2000} \\
               &                & E2 & $1.16[-9]$  & 0.3945(22) \cite{Safronova2017} & 0.3908(16) \cite{Letchumanan2005} \\
               & $5S_{1/2}$ & E2 & 2.53467 & 0.403(3) \cite{Kaur2021} & 0.372(25) \cite{Madej1990} \\[0.3ex]
\multicolumn{6}{l}{\underline{Ba$^+$ ion}} \\ 
$5D_{3/2}$ & $6S_{1/2}$ & M1 & $0.08[-6]$ & 81.0(5)$^*$ & 79.8(4.6) \cite{Yu1997} \\
               &                & E2 & 0.01234 & 81.4(1.4) \cite{Safronova2017} & 89(16) \cite{Gurell2007} \\
               &                & &  & 83(1) \cite{Kaur2021} &  \\ [1ex]
$5D_{5/2}$ & $5D_{3/2}$ & M1 & 0.00558 & 30.2(2)$^*$ & 32.0(4.6) \cite{Gurell2007}\\
               &                & E2 & $2.82[-7]$ & 30.34(48) \cite{Safronova2017} & 34.5(3.5) \cite{Madej1990-2} \\
               & $6S_{1/2}$ & E2 & 0.02754 & 30.81(65) \cite{Kaur2021} & \\
\hline \hline
\end{tabular}
\label{tab4}
\end{table}

Our calculated results for the energies of various states of the alkaline-earth metal ions are presented in Table \ref{tab1}. We find that in all the states across all the considered ions, correlation effects improve the results significantly over the DHF results and inclusion of triples contributions decrease the values by 0.2-0.5\% from the RCCSD method compared to the overall correlation contributions, which are about 4-7\% in the ground states and about 9-12\% in the metastable states. $+$Basis and $+$Breit contributions are on the order of 0.1\%, whereas the $+$VP corrections are generally found to be very small. Comparison of our final results with the experimental energies from the National Institute of Standards and Technology (NIST) database \cite{NIST} show small percentage differences. The calculated values can be improved further after inclusion of correlation effects from the quadruple excitations and self-energy corrections, which we defer to our future work but the present level of accuracy is sufficient enough to estimate properties of our interest using the calculated wave functions to the intended accuracy.

In the presence of external DC electric field $\vec{\cal E}$, the Stark shift due to the second-order correction to $E_v$ of the $|J_v, M_v \rangle$ state is $E_v^{(2)} = -\frac{1}{2} \alpha_d(J_v, M_v) |\vec{\cal E}|^2$. Conventionally, the dependency of $M_v$ is taken care by expressing $\alpha_d(J_v, M_v) = \alpha_d^S (J_v) + \frac{3 M_v^2 - J_v (J_v+1)}{J_v (2J_v -1)} \alpha_d^T(J_v)$, where $\alpha_d^S$ and $\alpha_d^T$ are referred as scalar and tensor components of $\alpha_d$. Detailed computational procedures for evaluating $\alpha_d^S$ and $\alpha_d^T$ in linear response (LR) RCC method, are given in Refs. \cite{Chakraborty2025, Chakraborty2025-2} (also refer to Supplemental Materials \cite{suppl}). In the bottom part of Table \ref{tab1}, we report our results for $\alpha_d^S$. In these quantities, the $+$Breit corrections are found to be consistently larger than the $+$VP contributions, and also outweighs the $+$Basis contributions. We also compare our final values with the available measurements \cite{Snow2007, Gallagher1982} and other precise estimations \cite{Safronova2011, Sahoo2009-2, Sahoo2009-3, Tchoukova2008, Jiang2009} in the same table. In Table \ref{tab2}, we present $\alpha_d^T$ values obtained from the DHF and RCC methods which show similar behavior as for $\alpha_d^S$. In fact, we provide additional values from the MP methods based on the Rayleigh-Schr\"odinger (MP$^{RS}$) and Brillouin-Wigner (MP$^{BW}$) approaches, and random phase approximation (RPA) in the Supplemental Materials \cite{suppl}. Due to unavailability of the experimental results, we compare our final values with other high-accuracy calculations from Refs. \cite{Safronova2011, Sahoo2009-2, Sahoo2009-3} and they seem to have good agreement overall. Apart from theoretical estimations in Refs. \cite{Sahoo2009-2, Sahoo2009-3}, other calculations mentioned above make use of experimental energies and/or E1 matrix elements to offer the precise values. Very good agreement between our {\it ab initio} results with the semi-empirical and measured values demonstrate potential of our employed LR-RCC method. It is to be noted that contributions from the triples are included in the uncertainties listed in the table, but they are discussed later. 

\begin{table*}[t!]
\setlength{\tabcolsep}{1pt}
\caption{Calculated $A_{hf}$ and $B_{hf}/Q_n$ values of the ground and metastable states of $^{43}$Ca$^+$, $^{87}$Sr$^+$, and $^{137}$Ba$^+$ using the DHF and many-body methods, and their comparisons with the literature values. The extracted $Q_n$ values of the above isotopes from precise measurements of $B_{hf}$ are also given.}
\centering
\begin{tabular}{l ccc ccc ccc}
\hline\hline
 & \multicolumn{3}{c}{Ca$^+$} & \multicolumn{3}{c}{Sr$^+$}  & \multicolumn{3}{c}{Ba$^+$}\\
\cline{2-4}\cline{5-7} \cline{8-10} \\
Method & $4s ~ ^2S_{1/2}$ & $3d ~ ^2D_{3/2}$ & $3d ~ ^2D_{5/2}$ & $5s ~ ^2S_{1/2}$ & $4d ~ ^2D_{3/2}$ & $4d ~ ^2D_{5/2}$ & $6s ~ ^2S_{1/2}$ & $5d ~ ^2D_{3/2}$ & $5d ~ ^2D_{5/2}$ \\
\hline 
\multicolumn{10}{c}{$A_{hf}$ values (in MHz)} \\
DHF  & $-588.26$ & $-33.21$ & $-14.15$ & $-734.39$  & $-31.14$ & $-12.98$ & 2935.35 & 128.41 & 51.5\\
RCCSD  & $-810.77$ & $-47.55$ & $-4.32$ & $-1009.14$ & $-46.15$ & 0.99 & 4094.24 & 189.74 & $-5.99$ \\
RCCSDT & $-805.41$ & $-47.60$ & $-3.25$ & $-997.32$ & $-46.26$ & 2.64 & 4026.31 & 191.21 & $-14.06$ \\
$+$Basis & $-0.34$ & $-0.06$ & $-0.03$ & $-0.83$ & $-0.10$ & $-0.05$ & 8.14 & 0.65 & 0.29 \\
$+$Breit & $-0.72$ & $-0.12$ & $-0.04$ & $-1.55$ & $-0.22$ & $-0.06$ & 6.70 & 1.19 & 0.36 \\
$+$VP  & $-0.38$ & $\sim 0.0$ & 0.01 & $-1.12$ & $-0.01$ & 0.02 & 8.67 & 0.11 & $-0.14$ \\
$+$BW & 0.88 & 0.02 & $-0.01$ & 3.15 & 0.05 & $-0.05$ & $-36.46$ & $-0.58$ & 0.60 \\[1ex]
Final & $-806(3)$ & $-47.8(3)$ & $-3.3(4)$ & $-998(4)$ & $-46.5(2)$ & 2.5(2) & 4013(9) & 193(4) & $-12.9(9)$ \\[0.5ex]
Theory \cite{Sahoo2009CaSrBa} & $-806.4(2.5)$ & $-47.3(3)$ & $-3.6(3)$ & $-1001.203$ & $-45.604$ & 2.145 &  & 190.89 & $-11.99$ \\
Theory \cite{Mani2010} & $-808.126$ & $-45.294$ & $-4.008$ & $-990.638$ & $-44.320$ & 2.168 & 4021.721 & 185.013 & $-12.593$ \\
Theory \cite{Itano2006} &  & $-47.27$ & $-4.84$ &  & $-45.60$ & $-2.77$ & & 192.99 & $-9.39$ \\
Expt. &  $-806.40207^a$ & $-48.3(1.6)^c$ & $-3.8(6)^d$  & $-1000.5(1.0)^f$  &  & 2.1743(14)$^h$ & 4018.2$^i$ & 189.7296(7)$^j$ & $-12.028(11)^j$ \\
 &  $-805(2)^b$ & $47.3(2)^d$  & $-3.8931(2)^e$  & $-1000.47367^g$ &   & &  & & \\
\hline 
\multicolumn{10}{c}{$B_{hf}/Q_n$ values (in MHz/b)} \\[0.5ex]
DHF   &  & 54.45 & 77.15 &   & 80.35 & 110.04  &  & 133.57 & 171.09 \\
RCCSD  &  & 68.63 & 95.90 &   & 116.48 & 160.16 &  & 194.51 & 256.34 \\
RCCSDT &  & 69.38 & 95.83 &   & 115.47 & 157.87 &  & 191.61 & 251.23 \\
$+$Basis &  & 0.10 & 0.13 &  & 0.26 & 0.35 &  & 0.53 & 0.66 \\
$+$Breit &  & 0.01 & 0.10  &   & $-0.06$ & 0.10 &  & $-0.30$ & $-0.07$ \\
$+$VP  &  & $-0.01$ & $\sim 0.0$ &  & $-0.01$ & $\sim 0.0$ &  & $-0.02$ & $-0.01$ \\[2ex]
Final &  & 69.5(8) & 96.1(9) &   & 115.7(8) & 158.3(10) &  & 191.8(20) & 251.8(25) \\
\hline
\multicolumn{10}{c}{$B_{hf}$ values (in MHz)} \\[0.5ex]
Expt. &  &  $-3.7(1.9)^d$ & $-3.9(6.0)^d$ &   &  & 49.11(6)$^h$ &  & 44.5408(17)$^j$ & 59.533(43)$^j$  \\
&  &  & $-4.241(4)^e$ &   &  &  &  &  &  \\
\hline
\multicolumn{10}{c}{$Q_n$ values (in b)} \\[0.5ex]
This work &  & \multicolumn{2}{c}{$-0.0441(4)$} &   & \multicolumn{2}{c}{0.3102(19)} &  & \multicolumn{2}{c}{0.2344(17)}  \\[0.2ex]
Other work &  &  \multicolumn{2}{c}{$-0.0444(6)$ \cite{Sahoo2009CaSrBa}}  &   &  \multicolumn{2}{c}{0.305(2) \cite{Sahoo2009CaSrBa}}   &  &  \multicolumn{2}{c}{0.246(1) \cite{Sahoo2009CaSrBa}}  \\
&  &  \multicolumn{2}{c}{$-0.0479(6)$ \cite{Tangarxiv}}  &   &  \multicolumn{2}{c}{0.323(20)\cite{Yu2004}}   &  &  \multicolumn{2}{c}{0.245(4) \cite{Pyykko2001}}  \\
&  &  \multicolumn{2}{c}{$-0.0408(8)$ \cite{Pyykko2001}}  &   &  \multicolumn{2}{c}{0.335(20)\cite{Pyykko2001}}   &  &  \multicolumn{2}{c}{}  \\
\hline\hline\\
\end{tabular}
Ref: $^a$\cite{Arbes1994}; $^b$\cite{Silverans1991}; $^c$\cite{Kurth1995}; $^d$\cite{Nortershauer1998}; $^e$\cite{Benhelm2007}; $^f$\cite{Buchinger1990}; $^g$\cite{Sunaoshi1993}; $^h$\cite{Barwood2003}; $^i$\cite{Becker1968}; $^j$\cite{Silverans1986}
\label{tab5}
\end{table*}

In Table \ref{tab2}, we report electric quadrupole (E2) moments of the metastable clock states of the considered ions and compare them with the literature values \cite{Roos2006, Barwood2004, Shaniv2016, Jiang2008, Itano2006}. Though our calculations and measured values are close, yet they are outside the quoted error bars. Compared to other calculations, our results are more precise and do not sometime agree with others due to inclusion of triples contributions. More importantly, analysis of MP, RPA and $+$Basis results, in the Supplemental Materials \cite{suppl}, suggests that the calculations reported using the B-spline basis functions in Ref. \cite{Jiang2008} and GTOs in the present case show very different trends of contributions from orbitals belonging to different $l$-quantum number. For example, Ref. \cite{Jiang2008} reported a large contribution from basis extrapolation with $l \ge 7$, amounting to nearly 0.9\% in some cases. By contrast, we find that the basis extrapolation effects are generally smaller than our quoted uncertainties. It may be attributed to the way continuum are generated in both the works.

Inclusion of triples effects in the evaluation of $\alpha_d^S$ and $\alpha_d^T$ is strenuous in the LR-RCC method and it requires to determine amplitudes for both the unperturbed and perturbed wave operators. $\vec{\bf D}$ being a vector operator, the perturbed RCC operators generates unusually large number of amplitudes, especially for the metastable $D$ states, compared to their unperturbed counterparts at the given set of active orbitals. This is why we only estimated triples contributions to the ground state of $\alpha_d$. However to gauge their influences to the $\alpha_d$ values of the $D$ states, we also estimated them in the lighter Ca$^+$ ion. From Table \ref{tab12}, we find that triples contributions to $\alpha_d$ are smaller than 0.5\% in the ground states and 1.5\% in the $D$ states. To understand the reason for small contributions arising from triples to the ground state $\alpha_d$ values, we analyzed changes in the excitation energies and E1 matrix elements. We find that when moving from RCCSD to RCCSDT, the values of the E1 matrix elements decrease, from 2.912 atomic units (a.u.) and 4.119 a.u. to 2.903 a.u. and 4.109 a.u., and at the same time energies also decrease, from 25419.02 cm$^{-1}$ and 25509.06 cm$^{-1}$ to 25145.23 cm$^{-1}$ and 25374.59 cm$^{-1}$, for the $4S_{1/2} \leftrightarrow 4P_{1/2}$ and $4S_{1/2}\leftrightarrow 4P_{3/2}$ transitions, respectively, in Ca$^+$. Since polarizability depends on the square of the E1 matrix elements and inversely on the transition energies, similar correlation trends in both the properties largely compensate to produce almost similar $\alpha_d$ values in the RCCSD and RCCSDT methods.

We now intend to fathom the observed inconsistencies in the lifetime studies, both from theoretical and experimental, of the metastable states of the considered clock candidates that are mainly governed by the forbidden magnetic dipole (M1) and E2 transition amplitudes. By evaluating the M1 ($A_{n\rightarrow k}^{M1}$) and E2 ($A_{n\rightarrow k}^{E2}$) transition rates (in s$^{-1}$) for the $|J_n, M_n \rangle \rightarrow |J_k, M_k\rangle$ transition, its lifetime is determined as $\tau_n = \frac{1}{\sum_{k,o} A_{n\rightarrow k}^o}$,
where the sum for $o$ runs over all possible decay modes (E2 and M1) and sum for $k$ runs for all possible decay states. We present our calculated E2 and M1 matrix elements from the DHF, RCCSD and RCCSDT methods along with other corrections in Table \ref{tab3}. These values from the MP methods and RPA can be found in the Supplemental Materials \cite{suppl}. Comparing our final results with other calculations \cite{Kaur2021, Safronova2017}, we find the E2 and M1 values are improved substantially due to inclusion of contributions from the triple excitations. Using the above discussed E2 and M1 matrix elements and wavelength values from the experiments \cite{NIST}, we present lifetimes for the metastable $^2D_{3/2}$ and $^2D_{5/2}$ states of the considered ions in Table \ref{tab4}. For completeness, we give the calculated transition probabilities from both the metastable states of each ion due to the E2 and M1 decay channels in the above table. Our analysis shows that, for all ions, the lifetimes are governed primarily by the E2 channel with an exception arises in the case of the $5D_{5/2}$ state of Ba$^+$, where the $5d~^2D_{5/2}\rightarrow 5d~^2D_{3/2}$ M1 channel provides a significant additional contribution. It can be observed from the above table that our $\tau$ values agree well with some of the calculations and measurements \cite{Safronova2017, Barton2000, Biemont2000, Mannervik1999, Letchumanan2005, Yu1997, Gurell2007, Madej1990}, yet differ significantly from others \cite{Kreuter2005, Madej1990-2, Kaur2021}. Since we use the E2 and M1 matrix elements that accounts more physical effects through the triples, $+$Breit and $+$VP compared to the earlier calculations, our estimated values are considered to be very reliable.

In Table \ref{tab5}, we report our calculated magnetic dipole hyperfine structure constants ($A_{hf}$) and ratio of $Q_n$ and electric quadrupole hyperfine structure constants ($B_{hf}/Q_n$) of the $^{43}$Ca$^+$, $^{87}$Sr$^+$ and $^{137}$Ba$^+$ ions from the DHF, RCCSD and RCCSDT methods including $+$Basis, $+$Breit, $+$VP and Bohr-Weisskopf effect ($+$BW) contributions. We find, our $A_{hf}$ values are in good agreement with other calculations and available precise measurements \cite{Arbes1994, Silverans1991, Kurth1995, Nortershauer1998, Benhelm2007, Buchinger1990, Sunaoshi1993, Barwood2003, Becker1968, Silverans1986, Sahoo2009CaSrBa, Itano2006, Mani2010}. Details of correlation trends through MP and RPA results, and also the nuclear magnetic moments used for estimating $A_{hf}$ by us are discussed in the Supplemental Materials \cite{suppl}. From our $A_{hf}$ values, we anticipate that the obtained $B_{hf}/Q_n$ are equally accurate. Combining our $B/Q_n$ calculations with the precisely measured $B_{hf}$ values from Refs. \cite{Benhelm2007, Barwood2003, Silverans1986}, we have extracted the $Q_n$ values of $^{43}$Ca, $^{87}$Sr and $^{137}$Ba. They are compared with the earlier extracted values \cite{Sahoo2009CaSrBa, Yu2004, Tangarxiv, Pyykko2001} in the above table. We find differences of about 4–9\% from previously reported values compared to ours. Since our calculations incorporate correlations and relativistic effects in a more rigorous manner, the inferred $Q_n$ values in the present work are expected to be more accurate.

This work is supported by ANRF with grant no. CRG/2023/002558 and Department of Space, Government of India. All calculations were performed on the ParamVikram-1000 HPC cluster at the Physical Research Laboratory (PRL), Ahmedabad, Gujarat, India.

\newpage
\clearpage

\onecolumngrid
\begin{center}
\textbf{\large Supplemental Materials: \\
Investigating Roles of Triple Excitations for High-precision Determination of Clock Properties of Alkaline Earth Metal Singly Charged Ions}
\end{center}
\twocolumngrid

For an atomic state $|J_vM_v\rangle$, $\alpha_d(J_v,M_v)$ depends on $J_v$ and $M_v$. Using the $M_v$-dependent factors, it yields \cite{Manakov1986, Stalnaker2006}
\begin{eqnarray}
\alpha_d(J_v,M_v)=\alpha_d^{S}(J_v) + \frac{3M_v^2-J_v(J_n+1)}{J_v(2J_v-1)}\alpha_d^{T}(J_v). 
\end{eqnarray}
Here $\alpha_d^{S}(J_v)$ and $\alpha_d^{T}(J_v)$ are called as the scalar and tensor polarizabilities, respectively. These $M_v$ independent quantities can be written in terms of reduced matrix elements as \cite{Manakov1986, Kozlov1999}
\begin{eqnarray}
 \alpha_d^S(J_v) &=& C_0\sum_{k}\frac{|\langle J_v||D||J_k \rangle|^2}{E^{(0)}_{J_v} - E^{(0)}_{J_k}}
 \label{eqas}
\end{eqnarray}
and
\begin{eqnarray}
 \alpha_d^T(J_v)&=&\sqrt{\frac{40J_v(2J_v-1)}{3(J_v+1)(2J_v+3)(2J_v+1)}}\sum_k (-1)^{J_v+J_k+1}\nonumber\\
 &&\times\left\{ \begin{array}{ccc}
                J_v& 2 & J_v\\
                1 & J_k &1 
 \end{array}\right\} (-1)^{J_v-J_k} \frac{|\langle J_v||D||J_k \rangle|^2}{E^{(0)}_{J_v} - E^{(0)}_{J_k}} \nonumber\\
 &=& \sum_{k} C_k \frac{|\langle J_n|| D||J_k \rangle|^2}{E^{(0)}_{J_v} - E^{(0)}_{J_k}},  
 \label{eqat}
\end{eqnarray}
where $C_0 =  - \frac{2}{3(2J_v+1)}$, $C_k = \sqrt{\frac{40J_n(2J_v-1)}{3(J_v+1)(2J_v+3)(2J_v+1)}}  \times  (-1)^{J_v+J_k+1} \left\{ \begin{array}{ccc}
                    J_v& 2 & J_v\\
                  1 & J_k &1 
 \end{array}\right\} $
with $D=\sum_q d_q$ is the E1 operator. In the linear response (LR) approach, we express $\alpha_d^{S/T}$ as
\begin{eqnarray}
\alpha_d^{S/T}&=&\langle \Psi_v^{(0)}|\tilde{D}^{S/T}|\Psi_v^{(1)} \rangle+\langle \Psi_v^{(1)}|\tilde{D}^{S/T}|\Psi_v^{(0)} \rangle\nonumber\\
 &=&2\langle \Psi_v^{(0)}|\tilde{D}^{S/T}|\Psi_v^{(1)} \rangle ,
\end{eqnarray}
where $|\Psi_v^{(0)} \rangle$ and $|\Psi_v^{(1)} \rangle$ denote the unperturbed wave function of the atomic Hamiltonian ($H_{at}$) and first-order perturbed wave function due to the E1 interaction, respectively. The term $\tilde{D}^{S/T}$ refers to the effective dipole operators for scalar and tensor components, given by $\tilde{D}^S = C_0 D$ and $\tilde{D}^T = \sum_k C_k D$.

To begin with, we consider $H_{at}$ at the Dirac-Coulomb (DC) approximation, given by (in a.u.)
\begin{eqnarray}
H_{at} &\equiv& \sum_i \left [c {\vec \alpha}_i^D \cdot {\vec p}_i+(\beta_i^D-1)c^2+V_n(r_i)\right ] +\sum_{i,j>i}\frac{1}{r_{ij}}. \nonumber 
\end{eqnarray}
Here, $\alpha^D$ and $\beta^D$ are the Dirac matrices, $\vec{p}$ is the single-particle momentum operator, $V_n(r)$ represents the nuclear potential felt by an electron, and $\frac{1}{r_{ij}}$ represents the Coulomb repulsion between two electrons. Since all the considered atomic states have closed-core and a valence orbital with different parity and angular momentum, we consider the $V^{N-1}$ potential formalism, where $N$ is the total number of electron of the considered singly charged ion, to produce the initial wave function, $|\Phi_0 \rangle$, using the DHF Hamiltonian, $H_{DHF}$. 

\begin{table*}[t!]
\setlength{\tabcolsep}{7pt}
\caption{Calculated energies (in cm$^{-1}$) and $\alpha_d^S$ values (in a.u.) of the clock states of Ca$^+$, Sr$^+$, and Ba$^+$ using the DHF and various many-body methods.}
\centering
\begin{tabular}{l ccc ccc ccc}
\hline\hline
 & \multicolumn{3}{c}{Ca$^+$} & \multicolumn{3}{c}{Sr$^+$}  & \multicolumn{3}{c}{Ba$^+$}\\
\cline{2-4}\cline{5-7} \cline{8-10} \\
Method & $4s ~ ^2S_{1/2}$ & $3d ~ ^2D_{3/2}$ & $3d ~ ^2D_{5/2}$ & $5s ~ ^2S_{1/2}$ & $4d ~ ^2D_{3/2}$ & $4d ~ ^2D_{5/2}$ & $6s ~ ^2S_{1/2}$ & $5d ~ ^2D_{3/2}$ & $5d ~ ^2D_{5/2}$ \\
\hline \\
\multicolumn{10}{c}{Energies in cm$^{-1}$} \\
DHF    & 91440.02 & 72618.65 & 72594.55 &  84041.96 & 67384.80 & 67241.97 & 75339.35 & 68139.75 & 67665.52 \\
MP(2)  & 96176.40 & 82729.13 & 82650.14 & 89577.17 & 75061.22 & 74741.76 & 81794.94 & 76934.16 & 76014.21 \\
RCCSD  & 95906.31 & 82293.80 & 82212.33 & 89094.93  & 74404.68 & 74101.58 & 80883.80 & 75738.94 & 74907.46 \\
RCCSDT & 95681.86 & 81912.20 & 81836.51 & 88859.83 & 74239.34 & 73953.86 & 80482.23 & 75420.56 & 74629.95  \\
\hline \\
\multicolumn{10}{c}{$\alpha_d^S$ values in a.u.} \\
DHF    & 98.90 & 104.92 & 102.24 & 127.03 & 148.90 & 139.42 & 185.52 & 91.29 & 88.78 \\
MP(2)$^{RS}$  & 86.49 & 93.14 & 90.85 & 106.46 & 126.86 & 120.02 & 146.377 & 71.38 & 70.49 \\
MP(3)$^{RS}$  & 90.91 & 104.88 & 100.19 & 113.17 & 163.90 & 145.56 & 162.94 & 76.30 & 72.89 \\
MP(2)$^{BW}$  & 72.56 & 36.64 & 36.55 & 86.21 & 61.29 & 61.09 & 112.04 & 49.52 & 50.59 \\
MP(3)$^{BW}$  & 75.31 & 30.33 & 30.20 & 90.17 & 60.20 & 59.41 & 121.09  & 49.95 & 50.50 \\
RPA    & 93.28 & 99.78 & 97.27 & 117.80 & 139.15 & 130.53 & 168.37 & 82.07 & 80.17 \\
RCCSD  & 76.33  & 32.01 & 31.87 & 92.05 & 64.53 & 63.32 & 125.37 & 52.45 & 52.56  \\
\hline\hline\\
\end{tabular}
\label{tabs1}
\end{table*}

\begin{table*}[t]
\setlength{\tabcolsep}{5pt}
\caption{$\alpha_d^{T}$ and $\Theta$ values (both are in a.u.) of the metastable $^2D_{3/2,5/2}$ states of Ca$^{+}$, Sr$^{+}$, and Ba$^{+}$ using the DHF and different many-body methods.}
\centering
\begin{tabular}{l cc cc cc cc cc cc}
\hline\hline
 & \multicolumn{4}{c}{Ca$^+$} & \multicolumn{4}{c}{Sr$^+$}  & \multicolumn{4}{c}{Ba$^+$}\\
\cline{2-5}\cline{6-9} \cline{10-13} \\
Method & \multicolumn{2}{c}{$3D_{3/2}$} & \multicolumn{2}{c}{$3D_{5/2}$} & \multicolumn{2}{c}{$4D_{3/2}$} & \multicolumn{2}{c}{$4D_{5/2}$} & \multicolumn{2}{c}{$5D_{3/2}$} & \multicolumn{2}{c}{$5D_{5/2}$} \\
   & $\alpha_d^T$ & $\Theta$ & $\alpha_d^T$ & $\Theta$ & $\alpha_d^T$ & $\Theta$ & $\alpha_d^T$ & $\Theta$ & $\alpha_d^T$ & $\Theta$ & $\alpha_d^T$ & $\Theta$ \\
\hline \\
DHF & $-66.39$ & 1.712 & $-90.68$ & 2.451 &  $-93.83$ & 2.469 &  $-118.32$ & 3.559 & $-45.50$ & 2.733 & $-58.63$ & 3.994 \\
MP(2)$^{RS}$  & $-58.79$ & 1.666 & $-80.41$ & 2.386 & $-79.50$ & 2.390 & $-100.93$ & 3.457 & $-34.97$ & 2.595 & $-45.97$ & 3.809 \\
MP(3)$^{RS}$  & $-70.27$ &  1.152 & $-93.04$ & 1.653  & $-112.12$ & 1.916 &  $-130.32$ & 2.780 & $-40.08$ & 2.153 & $-48.84$ & 3.185 \\
MP(2)$^{BW}$  & $-19.24$ & 1.666 & $-27.21$ &  2.386 & $-31.99$ & 2.396 & $-44.33$ & 3.458 & $-20.15$ & 2.597 & $-28.12$ & 3.811 \\
MP(3)$^{BW}$  & $-16.29$ & 1.265 & $-22.89$ & 1.814 & $-33.03$ & 2.008 & $-44.72$ & 2.908 & $-20.88$ & 2.249 & $-28.22$ & 3.315 \\
RPA & $-63.15$ & 1.687 & $-86.29$ & 2.416 & $-87.71$  & 2.426 & $-110.87$ & 3.501 & $-41.22$ & 2.640 & $-53.43$ & 3.874 \\
RCCSD & $-17.05$ & 1.288 & $-23.99$ & 1.846 & $-35.72$ & 2.038 & $-48.02$ & 2.948 & $-22.45$ & 2.278 & $-30.14$ & 3.353 \\
RCCSDT & - & 1.296 & - & 1.857 & - & 2.037 & - & 2.984 & - & 2.279 & - & 3.354 \\
\hline\hline\\
\end{tabular}
\label{tabs2}
\end{table*}

At the first step, we express the exact unperturbed wave function of the closed-core, $|\Psi^{(0)}_0\rangle$, due to $H_{at}$ by
\begin{eqnarray}
|\Psi_0^{(0)}\rangle=\Omega_0^{(0)}|\Phi_0\rangle ,
\end{eqnarray}
where $\Omega_0^{(0)}$ is referred to as the wave operator, which accounts for the electron correlation effects arising from the residual interaction $V_{res}=H_{at}-H_{DHF}$. In order to obtain the desired wave function of an intended state  we append the required valence orbital, $v$, to the closed-core configuration in the next step by defining the modified DHF wave function as $|\Phi_v \rangle = a_v^{\dagger} |\Phi_0 \rangle$. This follows the final unperturbed wave function in the wave operator formalism 
\begin{eqnarray} 
|\Psi_v^{(0)}\rangle=(\Omega_0^{(0)}+\Omega_v^{(0)})|\Phi_v\rangle ,
\end{eqnarray} 
where $\Omega_v^{(0)}$ is responsible for accounting the correlation effects involving the electron from the valence orbital $v$. Similarly, the corresponding first-order perturbed wave functions can be expressed as
\begin{eqnarray}
|\Psi_0^{(1)}\rangle=\Omega_0^{(1)}|\Phi_0\rangle 
\end{eqnarray}
and
\begin{eqnarray} 
|\Psi_v^{(0)}\rangle=(\Omega_0^{(1)}+\Omega_v^{(1)})|\Phi_v\rangle ,
\end{eqnarray} 
where superscript (1) on wave operators stands for the first-order perturbation. In the DHF method contribution from $V_{res}$ is completely neglected; thus $\Omega_0^{(0)}=\Omega_v^{(0)}=1$. 

In order to understand roles correlation effects in different properties of our interest, we start calculations considering $V_{res}$ as perturbation in the second-order and third-order perturbation theory (denoted as MP(2) and MP(3), respectively), then include $V_{res}$ to all-orders through the random phase approximation (RPA) and relativistic coupled-cluster (RCC) theory. In the MP method, amplitudes of the wave operators are obtained using the Bloch equation \cite{Lindgren1986} 
\begin{eqnarray}
\bigg[\Omega^{(0)}_0,H_{DHF}\bigg]=\bigg(V_{res}\Omega^{(0)}_0 -\Omega^{(0)}_0 [V_{res} \Omega^{(0)}_0 ] \bigg)_{l}  
\end{eqnarray}
and 
\begin{eqnarray}\label{wav1}
\bigg[\Omega_v^{(0)}, H_{DHF}\bigg]&=&\bigg( V_{res} (\Omega^{(0)}_0+\Omega^{(0)}_v) \nonumber\\
&& -\Omega^{(0)}_v [V_{res} (\Omega^{(0)}_0 +\Omega^{(0)}_v) ] \bigg)_{l} , 
\end{eqnarray}
where `$l$' means that only the linked diagrams will contribute to the wave operator. It should be noted that energy of the state is given by
\begin{eqnarray}
 E_v=\langle\Phi_v|V_{res} (\Omega^{(0)}_0 +\Omega^{(0)}_v)|\Phi_v\rangle .  
\end{eqnarray}

In the Rayleigh-Schr\"odinger MP approach (MP$^{RS}$ method), the wave function and energies are evaluated order-by-order explicitly \cite{Lindgren1986, Bartlett2009}. On the other hand, in the Brillouin-Wigner MP approach (MP$^{BW}$ method), amplitudes of the wave operators are obtained by rewriting Eq. (\ref{wav1}) as \cite{Lindgren1986, Bartlett2009}
\begin{eqnarray}
\bigg[\Omega_v^{(0)}, H_{DHF}\bigg]&=&\bigg( V_{res} (\Omega^{(0)}_0+\Omega^{(0)}_v)\bigg)_{l} -\Omega^{(0)}_v E_v  .
\end{eqnarray}
Similarly, the first-order perturbed wave operators in the MP$^{RS/BW}$ method can be derived by extending the Bloch equation \cite{Lindgren1986, Bartlett2009, Chakraborty2023}
\begin{eqnarray}
\bigg[\Omega_0^{(1)}, H_{DHF}\bigg] &=& (D \Omega_0^{(0)} +V_{res} \Omega_0^{(1)} )_l  
\end{eqnarray}
and
\begin{eqnarray}
\bigg[\Omega_v^{(1)}, H_{DHF}\bigg] &=& \bigg(D (\Omega_0^{(0)} + \Omega_v^{(0)}) + V_{res} ( \Omega_0^{(1)} +\Omega_v^{(1)} )\bigg)_l \nonumber\\
&&  - \Omega_v^{(1)}E_v.  
\end{eqnarray}

\begin{table}[t!]
\setlength{\tabcolsep}{4pt}
\caption{Convergence of the MP(3)$^{RS}$ and MP(3)$^{BW}$ values for the estimated $\Theta$ (in a.u.) of the metastable states of Ca$^{+}$, Sr$^{+}$, and Ba$^{+}$ with different sizes of basis functions.}
\centering
\begin{tabular}{l cc c cc c cc }
\hline\hline
& \multicolumn{2}{c}{Ca$^+$} && \multicolumn{2}{c}{Sr$^+$}  && \multicolumn{2}{c}{Ba$^+$}\\
\cline{2-3}\cline{5-6} \cline{8-9}\\
Basis & $3D_{3/2}$ & $3D_{5/2}$ && $4D_{3/2}$ & $4D_{5/2}$ && $5D_{3/2}$ & $5D_{5/2}$ \\
\hline\\
\multicolumn{9}{c}{\underline{MP(3)$^{RS}$}}\\
$l_{max}=4$ & 1.186 & 1.701 && 1.961 & 2.844 && 2.203 & 3.256 \\[1ex]
$l_{max}=6$ & 1.152 & 1.653 && 1.916 & 2.780 && 2.153 & 3.185 \\[1ex]
$l_{max}=9$ & 1.147 & 1.647 && 1.909 & 2.770 && 2.144 & 3.171 \\[1ex]
\hline\\
\multicolumn{9}{c}{\underline{MP(3)$^{BW}$}}\\
$l_{max}=4$ & 1.272 & 1.845 && 2.040 & 2.954 && 2.284 & 3.365 \\[1ex]
$l_{max}=6$ & 1.265 & 1.814 && 2.008 & 2.908 && 2.249 & 3.315\\[1ex]
$l_{max}=9$ & 1.262 & 1.810 && 2.003 & 2.901 && 2.243 & 3.306 \\[1ex]
\hline\hline\\
\end{tabular}
\label{tabs-basis}
\end{table}

\begin{table*}[t!]
\setlength{\tabcolsep}{0.5pt}
\caption{Reduced E2 and M1 amplitudes (in a.u.) of the clock transitions in Ca$^+$, Sr$^+$, and Ba$^+$ ions with various many-body methods. Numbers appearing as $a[b]$ mean $a\times10^b$.}
\centering
\begin{tabular}{l ccc ccc ccc}
\hline\hline
 & \multicolumn{3}{c}{Ca$^+$} & \multicolumn{3}{c}{Sr$^+$}  & \multicolumn{3}{c}{Ba$^+$}\\
\cline{2-4}\cline{5-7} \cline{8-10} \\
Method & $S_{1/2}-D_{3/2}$ & $S_{1/2} - D_{5/2}$  & $D_{3/2} - D_{5/2}$ & $S_{1/2}- D_{3/2}$ & $S_{1/2} - D_{5/2}$  & $D_{3/2} - D_{5/2}$ & $S_{1/2} - D_{3/2}$ & $S_{1/2} - D_{5/2}$  & $D_{3/2} - D_{5/2}$ \\
\hline \\
\multicolumn{10}{c}{E2 amplitudes} \\
DHF    & 9.767 & 11.978 & 5.018 & 12.969 & 15.973 & 7.261 & 14.764 & 18.385 & 8.092 \\
MP(2)$^{RS}$  & 9.608 & 11.784 & 4.884 & 12.700 & 15.651 & 7.052 & 14.300 & 17.844 & 7.712 \\
MP(3)$^{RS}$  & 7.449 & 9.150 & 3.380 & 10.648 & 13.172 & 5.654 & 11.939 & 15.011 & 6.423 \\
MP(2)$^{BW}$  & 9.609 & 11.785  & 4.885  & 12.701 & 15.653 & 7.053 & 14.304 & 17.848 & 7.717 \\
MP(3)$^{BW}$  & 7.884 & 9.679  & 3.712  &  11.001 & 13.591 & 5.921 & 12.416  & 15.563 & 6.698 \\
RPA    & 9.712 & 11.912 & 4.945 & 12.854 & 15.841 & 7.142 & 14.539 & 18.142 & 7.849\\
RCCSD  & 7.914 &  9.716 & 3.777 & 11.142 & 13.762 & 6.005 & 12.662 & 15.850 & 6.777 \\
RCCSDT & 7.959 & 9.770 & 3.803 & 11.159  & 13.779 & 6.008 & 12.700 & 15.890 & 6.797 \\
\hline \\
\multicolumn{10}{c}{M1 amplitudes} \\
DHF    & $-4.1[-7]$ &  & 1.549 &  $1.6[-6]$ & & 1.549 & $5.5[-6]$ & & 1.549 \\
MP(2)$^{RS}$   & $3.8[-6]$ & & 1.528 & $2.7[-5]$ & & 1.528 & $1.5[-4]$ & & 1.520 \\
MP(3)$^{RS}$  & $9.9[-5]$ &  & 1.544 & $6.2[-5]$ &  & 1.545 & $3.3[-4]$ &  & 1.544 \\
MP(2)$^{BW}$  &  $3.8[-6]$  & & 1.528  & $2.7[-5]$ & & 1.528  & $1.4[-4]$  &   & 1.520 \\
MP(3)$^{BW}$  &  $9.9[-5]$  & & 1.544  & $6.1[-5]$ & & 1.544  &  $3.3[-4]$ &   & 1.542 \\
RPA   & $6.6[-6]$ & & 1.549 & $4.1[-5]$ &  & 1.549 & $2.3[-4]$ & & 1.550 \\
RCCSD  & $3.5[-4]$ &  & 1.555 & $3.0[-5]$ & & 1.553 & $3.1[-4]$ & & 1.554  \\
RCCSDT & $3.5[-4]$ &  & 1.555 & $3.2[-5]$ & & 1.553 &  $3.2[-4]$ &  & 1.554 \\
\hline\hline\\
\end{tabular}
\label{tabs3}
\end{table*} 

In the RPA, $\Omega^{(0)}_{0/v}$ are equivalent to the DHF method and the first-order perturbed wave operators are obtained by 
\begin{eqnarray}
(H_{DHF} - {\cal E}_0) \Omega_0^{(1)} |\Phi_0 \rangle =  - D | \Phi_0 \rangle - U_0^{RPA}  |\Phi_0 \rangle  
\end{eqnarray}
and
\begin{eqnarray}
(H_{DHF} - {\cal E}_v) \Omega_v^{(1) } |\Phi_v \rangle =  - D | \Phi_v \rangle - U_v^{RPA}  |\Phi_v \rangle ,
\end{eqnarray}
where ${\cal E}_0=\langle \Phi_0 | H_{DHF} | \Phi_0 \rangle$, ${\cal E}_v=\langle \Phi_v | H_{DHF} | \Phi_v \rangle$ and the RPA potentials are defined as
\begin{eqnarray}
  U_0^{RPA} | \Phi_0 \rangle = \sum_{a,b} \left [ \langle b | V_{res} \Omega_0^{(1)} |b \rangle |a \rangle  -  \langle b | V_{res} \Omega_0^{(1)}  |a \rangle |b \rangle \right. \nonumber  \\ 
  \left. + \langle b | \Omega_0^{(1)\dagger} V_{res}  | b \rangle |a \rangle -  \langle b |\Omega_0^{(1)\dagger} V_{res}  |a \rangle |b \rangle  \right ]   \nonumber\\ 
\end{eqnarray}
and
\begin{eqnarray}
  U_v^{RPA} | \Phi_v \rangle = \sum_b \left [ \langle b | V_{res} \Omega_0^{(1)} |b \rangle |v \rangle   -  \langle b | V_{res} \Omega_v^{(1)}  |v \rangle |b \rangle \right. \nonumber  \\ 
  \left. + \langle b | \Omega_0^{(1)\dagger} V_{res}  | b \rangle |v \rangle  -  \langle b |\Omega_0^{(1)\dagger} V_{res}  |v \rangle |b \rangle  \right ] .  \nonumber\\
\end{eqnarray}

In the RCC method, we express the exact wave function $|\Psi_v \rangle$ of a one-valence atomic system (with valence orbital $v$) as $|\Psi_v \rangle = e^S |\Phi_v \rangle$, where $S$ is an wave operator that accounts for excitation amplitudes to all-orders of the residual interactions that are neglected at producing the mean-field wave function $|\Phi_v \rangle$. In the expanded form, it yields $S=S_1 + S_2 + \cdots + S_N$ with total number of electrons in the system $N$ and the subscript denotes level of excitation of electrons from core orbitals of $|\Phi_v \rangle$ to its virtuals; i.e. in the RCCSD method $S=S_1+S_2$ and in the RCCSDT method $S=S_1+S_2+S_3$. We adopt the Fock-space approach to express $S=T+S_v$, where $T$ is the excitation operator responsible for exciting only the core electrons and $S_v$ takes care of excitation of the valence electron coupling with the core-electrons. Due to presence of only one valence electron in the considered systems, the exact form of wave function yields $|\Psi_v \rangle = e^T \{ 1+S_v\} |\Phi_v \rangle$. The amplitudes of the $T$ and $S_v$ operators are obtained by solving
\begin{eqnarray}
\langle \Phi_v^* | [(H e^T)_l - E_v] S_v + (H e^T)_l|\Phi_v \rangle = 0,    
\end{eqnarray}
where subscript $l$ means it retains only linked terms and $|\Phi_v^* \rangle$ denotes for core and valence excitation determinants. In the {\it ab initio} approach, the energy is calculated as $E_v= \langle \Phi_v | (H e^T)_l \{ 1 + S_v \}|\Phi_v \rangle$. Since both amplitude and energy solving equations are interdependent, they are solved simultaneously. In the LR-RCC method, the $T$ and $S_v$ operators are expanded up to the zeroth- and first-order due to $\vec{\bf D} \cdot \vec{\cal E}$ for a small value of $|\vec{\cal E}|$; i.e. $T=T^{(0)}+ |\vec{\cal E}| T^{(1)}$ and $S_v=S_v^{(0)}+ |\vec{\cal E}| S_v^{(1)}$.

In the RCC theory {\it ansatz}, the unperturbed wave operators are defined as 
\begin{eqnarray}
\Omega_0^{(0)} = e^{T^{(0)}}  
\end{eqnarray}
and
\begin{eqnarray}
\Omega_v^{(0)} = e^{T^{(0)}} S_v^{(0)}.
\label{eqrcc}
\end{eqnarray}

Extending these definitions to the first-order perturbed wave functions, we can define the corresponding wave operators as 
\begin{eqnarray}
\Omega_0^{(1)} = e^{T^{(0)}} T^{(1)} 
\end{eqnarray}
and
\begin{eqnarray}
\Omega_v^{(1)} = e^{T^{(0)}} \left ( S_v^{(1)} + T^{(1)} S_v^{(0)} \right ).
\end{eqnarray}

Following the second-order energy expression $E_v^{(2)} = 2 \langle \Psi_v^{(0)} | \vec{\bf D} \cdot \vec{\cal E} | \Psi_v^{(1)} \rangle$, we arrive at the LR-RCC expressions
\begin{eqnarray}
\alpha_d^{S} &=& 
2\frac{\langle \Phi_v |\{ 1+ S_v^{(0)} \}^{\dagger} \bar{\tilde{D}}^S \{T^{(1)}(1+ S_v^{(0)}) + S_v^{(1)}\} |\Phi_v \rangle}{\langle \Phi_v | \{S_v^{(0)\dagger} +1 \} \bar{N} \{ 1+ S_v^{(0)} \} |\Phi_v \rangle} \nonumber \\  && + 2 \langle \Phi_0 | \bar{\tilde{D}}^S T^{(1)} |\Phi_0 \rangle 
\label{pols}
\end{eqnarray}
and
\begin{eqnarray}
\alpha_d^{T} &=&  2\frac{\langle \Phi_v |\{ 1+ S_v^{(0)} \}^{\dagger} \bar{\tilde{D}}^T \{T^{(1)}(1+ S_v^{(0)}) + S_v^{(1)}\} |\Phi_v \rangle}{\langle \Phi_v | \{S_v^{(0)\dagger} +1 \} \bar{N} \{ 1+ S_v^{(0)} \} |\Phi_v \rangle} , \nonumber \\ && 
\label{polt}
\end{eqnarray}
where $\bar{\tilde{D}}^{S/T}=e^{T^{(0)\dagger}}\tilde{D}^{S/T} e^{T^{(0)}}$, $\bar{N}=e^{T^{(0)\dagger}}e^{T^{(0)}}$.

In order to evaluate the M1 and E2 transition matrix elements, electric quadrupole moment ($\Theta$), magnetic dipole hyperfine structure constants ($A_{hf}$) and electric quadrupole hyperfine structure constants ($B_{hf}$), we use the following general expression
\begin{eqnarray}
 \langle O \rangle_{fi} &=& \frac{ \langle \Psi_f^{(0)} | O | \Psi_i^{(0)} \rangle} {\sqrt{\langle \Psi_f^{(0)} | \Psi_f^{(0)} \rangle \langle \Psi_i^{(0)} | \Psi_i^{(0)} \rangle}} \nonumber \\
    &=& \frac{\langle \Phi_f | (\Omega_f^{(0)\dagger} + \Omega_0^{(0)\dagger}) O (\Omega_0^{(0)} + \Omega_i^{(0)}) | \Phi_i \rangle}{\sqrt{N_f N_i}},
\end{eqnarray}
where $N_{k=i,f}= \langle \Phi_k | (\Omega_k^{(0)\dagger} + \Omega_0^{(0)\dagger}) (\Omega_0^{(0)} + \Omega_k^{(0)}) | \Phi_k \rangle$ with the respective operator $O$. In the evaluation of $\Theta$, $A_{hf}$ and $B_{hf}$ expressions, we set $| \Psi_f^{(0)} \rangle = |\Psi_i^{(0)} \rangle$. 

To account for the Bohr-Weisskopf (BW) effect \cite{Bohr1950}, we define the nuclear magnetization function ($F(r)$) in the Fermi nuclear charge distribution approximation as  
\begin{eqnarray}
 F(r) &=& \frac{f_{WS}}{\cal N} [ (r/c)^3 -3(a/c)(r/c)^2 R_1((c-r)/a) \nonumber \\ && + 6 (a/c)^2(r/c)R_2((c-r)/a) - 6 (a/c)^3 \nonumber \\ 
 && \times R_3((c-r)/a) +6(a/c)^3R_3(c/a) ]
 \label{magd}
\end{eqnarray}
for $r \le c$ and
\begin{eqnarray}
 F(r) &=& 1- \frac{1}{\cal N} [ 3 (a/c) (r/c)^2 R_1((r-c)/a) \nonumber \\ && + 6 (a/c)^2(r/c)R_2((r-c)/a)]
\end{eqnarray}
for $r>c$, where 
\begin{eqnarray}
 {\cal N} = 1 + (a/c)^2 \pi^2 + 6 (a/c)^3 R_3(c/a)
\end{eqnarray}
and 
\begin{eqnarray}
 R_k(x) = \sum_{n=1}^{\infty} (-1)^{n-1} \frac{e^{-nx}}{n^k} . 
\end{eqnarray}
In Eq. (\ref{magd}) $f_{WF}=1$ when Woods-Saxon potential correction is not taken into account, else it is estimated using following 
expressions after neglecting the spin-orbit interaction within the nucleus \cite{Shabaev1997, Volotka2008}
\begin{eqnarray}
 f_{WS} &=& 1- \left (\frac{3}{\mu_I} \right ) ln \left ( \frac{r}{c} \right ) \left [- \frac{2I-1}{8(I+1)}g_S + (I-1/2)g_L \right ]   \nonumber
\end{eqnarray}
for $I=L+\frac{1}{2}$ and
\begin{eqnarray}
 f_{WS} &=& 1- \left (\frac{3}{\mu_I} \right ) ln \left ( \frac{r}{c} \right ) \left [ \frac{2I+3}{8(I+1)}g_S + \frac{I(2I+3)}{2(I+1)}g_L \right ]   \nonumber
\end{eqnarray}
for $I=L-\frac{1}{2}$. 

\begin{table*}[t!]
\setlength{\tabcolsep}{7pt}
\caption{Calculated $A_{hf}$ and $B_{hf}/Q_n$ values of the ground and metastable states of $^{43}$Ca$^+$, $^{87}$Sr$^+$, and $^{137}$Ba$^+$ using the DHF and many-body methods.}
\centering
\begin{tabular}{l ccc ccc ccc}
\hline\hline
 & \multicolumn{3}{c}{Ca$^+$} & \multicolumn{3}{c}{Sr$^+$}  & \multicolumn{3}{c}{Ba$^+$}\\
\cline{2-4}\cline{5-7} \cline{8-10} \\
Method & $4s ~ ^2S_{1/2}$ & $3d ~ ^2D_{3/2}$ & $3d ~ ^2D_{5/2}$ & $5s ~ ^2S_{1/2}$ & $4d ~ ^2D_{3/2}$ & $4d ~ ^2D_{5/2}$ & $6s ~ ^2S_{1/2}$ & $5d ~ ^2D_{3/2}$ & $5d ~ ^2D_{5/2}$ \\
\hline \\
\multicolumn{10}{c}{$A_{hf}$ values (in MHz)} \\
DHF  & $-588.26$ & $-33.21$ & $-14.15$ & $-734.39$  & $-31.14$ & $-12.98$ & 2935.35 & 128.41 & 51.52 \\
MP(2)$^{RS}$ & $-683.49$ & $-32.62$ & $-1.02$ & $-843.21$ & $-31.41$ & 1.40 & 3349.47 & 131.65 & $-9.88$ \\
MP(3)$^{RS}$ & $-796.22$ & $-44.10$ & $-6.49$ & $-1016.75$ & $-40.43$ & $-2.79$ & 4249.28 & 167.05 &  6.39 \\
MP(2)$^{BW}$  & $-683.16$ & $-32.70$ & $-1.17$ & $-842.67$ & $-31.52$ & 1.21 & 3346.50 & 132.19 & $-8.72$ \\
MP(3)$^{BW}$  & $-789.87$ & $-42.59$ & $-5.95$ & $-1004.41$ & $-39.47$ & $-2.55$ & 4165.22 & 162.98 & 5.74 \\
RPA   & $-710.77$ & $-34.51$ & 5.93 & $-873.80$ & $-34.85$ & 11.77 & 3478.63 & 149.92 & $-56.69$ \\
RCCSD  & $-810.77$ & $-47.55$ & $-4.32$ & $-1009.14$ & $-46.15$ & 0.99 & 4094.24 & 189.74 & $-5.99$ \\
RCCSDT & $-805.41$ & $-47.60$ & $-3.25$ & $-997.32$ & $-46.26$ & 2.64 & 4026.31 & 191.21 & $-14.06$ \\
\hline \\
\multicolumn{10}{c}{$B_{hf}/Q_{n}$ values (in MHz/b)} \\[0.5ex]
DHF   &  & 54.45 & 77.15 &   & 80.35 & 110.04  &  & 133.57 & 171.09 \\
MP(2)$^{RS}$  &  & 52.70 & 75.02 &  & 97.89 & 136.92 &  & 170.47 & 229.94 \\
MP(3)$^{RS}$&  & 70.85 & 100.39 &  & 114.42 & 158.51 &  & 190.93
& 253.38 \\
MP(2)$^{BW}$  &  & 53.01 & 75.47 &  & 98.29 & 137.45  &  & 171.36 & 230.95 \\
MP(3)$^{BW}$  &  & 68.63 & 97.27 &   & 112.28 & 155.68 &  & 187.74 & 249.51 \\
RPA   &  & 47.84 & 68.24 &   & 86.42 & 121.55 &  & 146.95 & 200.25 \\
RCCSD  &  & 68.63 & 95.90 &   & 116.48 & 160.16 &  & 194.51 & 256.34 \\
RCCSDT &  & 69.38 & 95.83 &   & 115.47 & 157.87 &  & 191.61 & 251.23 \\
\hline\hline\\
\end{tabular}
\label{tab5}
\end{table*}

Our calculated energies and $\alpha_d^S$ values for the clock states of the alkaline-earth-metal ions, obtained using DHF and different many-body methods, are summarized in Table \ref{tabs1}. To analyze correlation effects in the energies, we present contributions from the DC Hamiltonian at the DHF and MP(2) levels, together with results from the RCC methods. The data show that core-polarization (CP) effects, which first appear at the MP(2) level, contribute as much as 5–12\% to the total energies. In contrast, the difference between the RCCSD and MP(2) results is only about 0.3–1.6\%, indicating all-order CP and pair-correlation (PC) effects play a relatively minor role in energy calculation. To illustrate the role of correlation effects to $\alpha_d^S$, we include perturbative results from both the $RS$ and $BW$ perturbation schemes. It is evident that the $BW$ approach provides values closer to the final results. This is expected, since unlike the $RS$ scheme, the $BW$ method accounts for correlation-corrected energies in the property evaluation, leading to improved accuracy. A comparison of DHF results with MP(2)$^{RS}$ and MP(2)$^{BW}$ shows that the lower-order CP effects generally reduce the $\alpha_d^s$ values. However, the correlation trend from MP(2) to MP(3) differs significantly between the two perturbation schemes. For the $RS$ method, the PC contribution, which appears at the MP(3) level, increases $\alpha_d^s$. In contrast, the $BW$ method shows an increase at MP(3) only for the ground state; for the $D_{3/2,5/2}$ states, MP(3)$^{BW}$ values are smaller than MP(2)$^{BW}$, except in the case of Ba$^+$, where the difference is negligible. We also find that the RPA results remain much closer to the DHF values compared to those from MP(2). Since RPA includes CP contributions from MP(2)$^{RS}$ to all orders, this close agreement suggests that there are cancellations between the lowest-order and higher-order CP effects in determining $\alpha_d^S$. Finally, we compare these results with those obtained using the RCCSD method. The RCCSD approach accounts for CP and PC effects to all orders and incorporates correlation in the state energies, thereby providing more reliable values. 

In Table \ref{tabs2}, we present the values of $\alpha_d^T$ and $\Theta$ obtained from DHF and various many-body methods. The overall correlation behavior is generally consistent with that observed for $\alpha_d^S$. Owing to the inclusion of correlation effects in the energy values, the results from the $BW$ scheme are more accurate than those from the $RS$ perturbation scheme. Similar to the case of $\alpha_d^S$, the lowest-order CP contribution reduces the magnitude of the tensor polarizability. However, in the $RS$ scheme, the value of $\alpha_d^T$ increases when going from MP(2)$^{RS}$ to MP(3)$^{RS}$, whereas in the $BW$ scheme, the inclusion of PC effects produces the opposite trend, further decreasing the magnitude. Similar to the case of $\alpha_d^S$, we find that the lowest-order CP effect reduces the $\Theta$ values. However, due to cancellations between lower-order and higher-order CP contributions, the RPA results remain much closer to DHF values than to those from MP(2)$^{RS}$ or MP(2)$^{BW}$. Unlike $\alpha_d^S$, where the correlation trends in $RS$ and $BW$ perturbation theories differ, in the case of $\Theta$, both schemes show a consistent behavior; the inclusion of PC effects at the MP(3) level leads to a further reduction in the $\Theta$ values. 

In order to illustrate how the value of $\Theta$ varies with the size of the basis functions, we present in Table~\ref{tabs-basis} the results obtained using the MP(3)$^{RS}$ and MP(3)$^{BW}$ methods. For this purpose, we have employed three different basis sets characterized by maximum angular quantum numbers $l_{\text{max}} = 4, 6,$ and $9$. As can be seen from Table~\ref{tabs-basis}, in the case of the $RS$ scheme, the difference between the results obtained with $l_{\text{max}}=4$ and $l_{\text{max}}=6$ is about 2–3\%, while the corresponding change between $l_{\text{max}}=6$ and $l_{\text{max}}=9$ reduces to only 0.4\%. In contrast, for the $BW$ scheme, the variations are even smaller; the difference between the MP(3)$^{BW}$ results with $l_{\text{max}}=4$ and $l_{\text{max}}=6$ still exceeds 1\%, but the change between $l_{\text{max}}=6$ and $l_{\text{max}}=9$ is less than 0.3\%. This trend indicates that the basis with $l_{\text{max}}=6$ already incorporates the dominant part of the correlation contributions. Taking into account both this observation and the practical limitations of our computational resources, we have therefore carried out the RCC calculations with $l_{\text{max}}=6$. Nevertheless, in order to ensure that the contributions from higher angular momentum orbitals are not neglected, their contributions have been explicitly included in the manuscript as the $+$Basis correction, evaluated within the $BW$ scheme, since this approach accounts for correlation effects rigorously in energies and matrix elements, in contrast to the $RS$ scheme.  

We present our calculated E2 and M1 matrix elements in Table \ref{tabs3}. For the E2 matrix elements, we observe a correlation trend similar to that found for the quadrupole moment $\Theta$, which is expected. In this case as well, the PC effects are far more significant than the CP contributions, bringing the MP(3) results much closer to our recommended RCC values. As shown in Table \ref{tabs3}, the M1 transition amplitudes are significantly smaller in magnitude compared to the corresponding E2 transitions in all the considered ions. Interestingly, the two cases we consider, namely the $S_{1/2}-D_{3/2}$ and $D_{3/2}-D_{5/2}$ transitions, display distinct correlation trends. For the $S_{1/2}-D_{3/2}$ transition, both CP and PC effects increase the M1 amplitude relative to the DHF value, often by one or even two orders of magnitude, with the PC contribution being more prominent than the CP effect. By contrast, in the $D_{3/2}-D_{5/2}$ transition, a comparison between MP(2) and MP(3) results reveals an opposite interplay; while CP effects tend to reduce the M1 amplitude, the PC contributions act in the opposite direction and increase it. Despite these competing trends, there is strong cancellation among the different correlation effects, which results in only small net changes in the M1 amplitudes across the many-body methods. 

In Table \ref{tab5}, we present our results for the $A_{hf}$ and $B_{hf}/Q_{n}$ values. For $A_{hf}$ calculation, we have used $g_I = -0.376469$ for $^{43}$Ca, $g_I = -0.2430229$ for $^{87}$Sr, and $g_I = 0.6249$ for $^{137}$Ba \cite{Stone}. Similar to the E2 and M1 amplitudes, the differences between the $RS$ and $BW$ results for $A_{hf}$ and $B_{hf}/Q_{n}$ at a given perturbative level are found to be small. For the ground state, our analysis shows that both CP and PC effects contribute in the same direction, increasing the magnitude of $A_{hf}$ value. However, for $D_{5/2}$, $A_{hf}$ value is found to be very sensitive to correlation effects. As can be seen from the MP(2) results in Table \ref{tab5}, the CP effects contribute up to 600-1500\% in $D_{5/2}$ states. The $D_{5/2}$ states also exhibit a striking sensitivity to PC effects, with contributions reaching as high as 200\%. Furthermore, we observe that triple excitations contribute only marginally to $A_{hf}$ in the $S_{1/2}$ and $D_{3/2}$ states, but play a dominant role in the $D_{5/2}$ states, where they account for more than 50\% of the total value. This unusually large contribution also hints at non-negligible contributions from higher-order excitations, such as quadrupole excitations. Interestingly, the correlation behavior in $B_{hf}/Q_{n}$ is quite different. In this case, the sensitivity to correlation effects is far less pronounced. Contributions from CP, PC, and even triple excitations remain relatively small across all states, including the $D_{5/2}$ levels. This contrast between $A_{hf}$ and $B_{hf}/Q_{n}$ demonstrates the property-dependent nature of correlation and emphasizes the need for a case-by-case treatment when evaluating the hyperfine structure constants.


\begin{thebibliography}{5}

\bibitem{Olmschenk2009}
S. Olmschenk, D. Hayes, D.N. Matsukevich, P. Maunz, D.L. Moehring, K.C. Younge, C. Monroe, Measurement of the lifetime of the $6p~^2P_{1/2}^o$ level of Yb$^+$, Phys. Rev. A {\bf 80}, 022502 (2009).

\bibitem{Sherman2008}
J. A. Sherman, A. Andalkar, W. Nagourney, E.N. Fortson, Measurement of light shifts at two off-resonant wavelengths in a single trapped Ba$^+$ ion and the determination of atomic dipole matrix elements, Phys. Rev. A {\bf 78}, 052514 (2008).

\bibitem{Sahoo2009-3}
B. K. Sahoo, R. G. E. Timmermans, B. P. Das, and D. Mukherjee, Comparative studies of dipole polarizabilities in Sr$^+$, Ba$^+$, and Ra$^+$ and their applications to optical clocks, Phys. Rev. A {\bf 80}, 062506 (2009).

\bibitem{Fortson1993}
N. Fortson, Possibility of measuring parity nonconservation with a single trapped atomic ion, Phys. Rev. Lett. {\bf 70}, 2383 (1993).

\bibitem{Wansbeek2008}
L. Wansbeek, B. K. Sahoo, R. G. E. Timmermans, K. Jungmann, B. P. Das, and D. Mukherjee, Atomic parity nonconservation in Ra$^+$,  Phys. Rev. A {\bf 78}, 050501(R) (2008).

\bibitem{Sahoo2007}
B. K. Sahoo, B. P. Das, R. K. Chaudhuri, and Debashis Mukherjee, Theoretical studies of the 6$s~^2S_{1/2} \rightarrow$ 5$d~^2D_{3/2}$ parity-nonconserving transition amplitude in Ba$^+$ and associated properties Phys. Rev. A {\bf 75}, 032507 (2007).

\bibitem{Sahoo2009}
B. K. Sahoo, L. W. Wansbeek, K. Jungmann, and R. G. E. Timmermans, Light shifts and electric dipole matrix elements in Ba$^+$ and Ra$^+$ Phys. Rev. A {\bf 79}, 052512 (2009).

\bibitem{Gossel2013}
G. H. Gossel, V. A. Dzuba, and V. V. Flambaum, Calculation of strongly forbidden M1 transitions and g-factor anomalies in atoms considered for parity-nonconservation measurements, Phys. Rev. A {\bf 88}, 034501 (2013).

\bibitem{Ruiz2016}
R.F. Garcia Ruiz, M. L. Bissell, K. Blaum, {\it et al.}, Unexpectedly large charge radii of neutron-rich calcium isotopes, Nat. Phys. {\bf 12}, 594 (2016).

\bibitem{Berengut2018}
Julian C. Berengut, Dmitry Budker, Cédric Delaunay, Victor V. Flambaum, Claudia Frugiuele, Elina Fuchs, Christophe Grojean, Roni Harnik, Roee Ozeri {\it et al.}, Probing new long-range interactions by isotope shift spectroscopy, Phys. Rev. Lett. {\bf 120}, 091801 (2018).

\bibitem{Dutta2022}
Tarun Dutta, Adrián Pérez-Salinas, Jasper Phua Sing Cheng, José Ignacio Latorre, and Manas Mukherjee, Single-qubit universal classifier implemented on an ion-trap quantum device, Phys. Rev. A {\bf 106}, 012411 (2022).

\bibitem{Clark2021}
Craig R. Clark, Holly N. Tinkey, Brian C. Sawyer, Adam M. Meier, Karl A. Burkhardt, Christopher M. Seck, Christopher M. Shappert, Nicholas D. Guise, Curtis E. Volin {\it et al.}, High-fidelity Bell-state preparation with $^{40}$Ca$^+$ optical qubits, Phys. Rev. Lett. {\bf 127}, 130505 (2021).

\bibitem{Moller2007}
Ditte Møller, Jens L. Sørensen, Jakob B. Thomsen, and Michael Drewsen, Efficient qubit detection using alkaline-earth-metal ions and a double stimulated Raman adiabatic process, Phys. Rev. A {\bf 76}, 062321 (2007).


\bibitem{Craik2025}
Diana P. L. Aude Craik, An entanglement protocol to measure atomic parity violation at sub 0.1\% precision, arXiv:2503.20003 (Unpublished).

\bibitem{Barton2000}
P. A. Barton, C. J. S. Donald, D. M. Lucas, D. A. Stevens, A. M. Steane, and D. N. Stacey, Measurement of the lifetime of the 3$d~^2D_{5/2}$ state in $^{40}$Ca$^+$,  Phys. Rev. A {\bf 62}, 032503 (2000).

\bibitem{Kreuter2005}
A. Kreuter, C. Becher, G. P. T. Lancaster, A. B. Mundt, C. Russo, H. Häffner, C. Roos, W. Hänsel, F. Schmidt-Kaler, R. Blatt {\it et al.}, Experimental and theoretical study of the 3$d~^2D$-level lifetimes of $^{40}$Ca$^+$, Phys. Rev. A {\bf 71}, 032504 (2005).

\bibitem{Mannervik1999}
S. Mannervik, J. Lidberg, L.-O. Norlin, P. Royen, A. Schmitt, W. Shi, and X. Tordoir, Lifetime measurement of the metastable 4$d~^2D_{3/2}$ level in Sr$^+$ by optical pumping of a stored ion beam,  Phys. Rev. Lett. {\bf 83}, 698 (1999).

\bibitem{Biemont2000}
E. Biémont, S. Mannervik, L.-O. Norlin, P. Royen, A. Schmitt, W. Shi, and X. Tordoir, Lifetimes of metastable states in Sr II, Eur. Phys. J. D {\bf 11}, 355 (2000).

\bibitem{Letchumanan2005}
V. Letchumanan, M. A. Wilson, P. Gill, and A. G. Sinclair, Lifetime measurement of the metastable 4$d~^2D_{5/2}$ state in $^{88}$Sr$^+$ using a single trapped ion, Phys. Rev. A {\bf 72}, 012509 (2005).

\bibitem{Yu1997}
N. Yu, W. Nagourney, and H. Dehmelt, Radiative lifetime measurement of the Ba$^+$ metastable $D_{3/2}$ state, Phys. Rev. Lett. {\bf 78}, 4898 (1997).

\bibitem{Gurell2007}
J. Gurell, E. Biémont, K. Blagoev, V. Fivet, P. Lundin, S. Mannervik, L.-O. Norlin, P. Quinet, D. Rostohar, P. Royen {\it et al.}, Laser-probing measurements and calculations of lifetimes of the 5$d~^2D_{3/2}$ and 5$d~^2D_{5/2}$ metastable levels in Ba II, Phys. Rev. A {\bf 75}, 052506 (2007).

\bibitem{Madej1990}
A. A. Madej and J. D. Sankey, Single, trapped Sr$^+$ atom: laser cooling and quantum jumps by means of the 4$d~^2D_{5/2}-$5$s~^2S_{1/2}$ transition, Opt. Lett. {\bf 15}, 634 (1990).

\bibitem{Madej1990-2}
A. A. Madej and J. D. Sankey, Quantum jumps and the single trapped barium ion: Determination of collisional quenching rates for the 5$d~^2D_{5/2}$ level, Phys. Rev. A {\bf 41}, 2621 (1990).

\bibitem{Arbes1994}
F. Arbes, M. Benzing, Th. Gudjons, F. Kurth, and G. Werth, Precise determination of the ground state hyperfine structure splitting of $^{43}$Ca II, Z. Phys. D: At., Mol. Clusters {\bf 31}, 27 (1994).

\bibitem{Silverans1991}
R. E. Silverans, L. Vermeeren, R. Neugart, and P. Lievens, Hyperfine structure constants of the Ca II states 4 $s~^2S_{1/2}$ and 4$p~^2 P_{1/2,3/2}$ and the nuclear quadrupole moment of $^{43}$Ca,  Z. Phys. D {\bf 18}, 351 (1991).

\bibitem{Kurth1995}
F. Kurth, T. Gudjons, B. Hiblert, T. Reisinger, G. Werth, and A-M. Martensson, Doppler free “dark resonances” for hyperfine measurements and isotope shifts in Ca$^{+}$ isotopes in a Paul trap, Z. Phys. D {\bf 34}, 227 (1995).

\bibitem{Nortershauer1998}
W. Nortershauer {\it et al.}, Isotope shifts and hyperfine structure in the transitions in calcium II, Eur. Phys. J. D {\bf 2}, 33 (1998).

\bibitem{Benhelm2007}
J. Benhelm, G. Kirchmair, U. Rapol, T. Korber, C. F. Roos, and R. Blatt, Measurement of the hyperfine structure of the $S_{1/2}-D_{5/2}$ transition in $^{43}$Ca$^+$, Phys. Rev. A {\bf 75}, 032506 (2007).

\bibitem{Buchinger1990}
F. Buchinger {\it et al.}, Systematics of nuclear ground state properties in $^{78-100}$Sr by laser spectroscopy,  Phys. Rev. C {\bf 41}, 2883 (1990).

\bibitem{Sunaoshi1993}
H. Sunaoshi {\it et al.}, A precision measurement of the hyperfine structure of $^{87}$Sr$^+$, Hyperfine Interact. {\bf 78}, 241 (1993).

\bibitem{Barwood2003}
G. P. Barwood, K. Gao, P. Gill, G. Huang, and H. A. Klein, Observation of the hyperfine structure of the $^2S_{1/2}- ^2D_{5/2}$ transition in $^{87}$Sr$^+$, Phys. Rev. A {\bf 67}, 013402 (2003).

\bibitem{Becker1968}
W. Becker, W. Fischer, and H. Huhnermann, The hyperfine structure of the Ba II-resonance lines and the quadrupole moments of the odd barium isotopes, Z. Phys. {\bf 216}, 142 (1968).

\bibitem{Silverans1986}
R. E. Silverans, G. Borghs, P. De Bisschop, and M. Van Hove, Hyperfine structure of the 5$d~^2D_J$ states in the alkaline-earth Ba ion by fast-ion-beam laser-rf spectroscopy, Phys. Rev. A {\bf 33}, 2117 (1986).

\bibitem{Snow2007}
E. L. Snow and S. R. Lundeen, Fine-structure measurements in high-L $n=$17 and 20 Rydberg states of barium, Phys. Rev. A {\bf 76}, 052505 (2007).

\bibitem{Gallagher1982}
T. F. Gallagher, R. Kachru, and N. H. Tran, Radio frequency resonance measurements of the Ba 6$sng$-6$snh$-6$sni$-6$snk$ intervals: An investigation of the nonadiabatic effects in core polarization, Phys. Rev. A {\bf 26}, 2611 (1982).

\bibitem{Roos2006}
C. F. Roos, M. Chwalla, K. Kim, M. Riebe, and R. Blatt, ‘Designer atoms’ for quantum metrology, Nature (London) {\bf 443}, 316 (2006).

\bibitem{Barwood2004}
G. P. Barwood, H. S. Margolis, G. Huang, P. Gill, and H. A. Klein, Measurement of the electric quadrupole moment of the 4$d~^2D_{5/2}$ level in $^{88}$Sr$^+$, Phys. Rev. Lett. {\bf 93}, 133001 (2004).

\bibitem{Shaniv2016}
R. Shaniv, N. Akerman, and R. Ozeri, Atomic quadrupole moment measurement using dynamic decoupling, Phys. Rev. Lett. {\bf 116}, 140801 (2016).

\bibitem{Jiang2008}
Dansha Jiang, B. Arora, and M. S. Safronova, Electric quadrupole moments of metastable states of Ca$^+$, Sr$^+$, and Ba$^+$, Phys. Rev. A {\bf 78}, 022514 (2008).

\bibitem{Itano2006}
W. M. Itano, Quadrupole moments and hyperfine constants of metastable states of Ca$^+$, Sr$^+$, Ba$^+$, Yb$^+$, Hg$^+$, and Au, Phys. Rev. A {\bf 73}, 022510 (2006).

\bibitem{Sur2006}
C. Sur, K. V. P. Latha, B. K. Sahoo, R. K. Chaudhuri, B. P. Das, and D. Mukherjee, Electric quadrupole moments of the $D$ states of alkaline-earth-metal ions, Phys. Rev. Lett. {\bf 96}, 193001 (2006).

\bibitem{Safronova2011}
M. S. Safronova, U.I. Safronova, Blackbody radiation shift, multipole polarizabilities, oscillator strengths, lifetimes, hyperfine constants, and excitation energies in Ca$^+$, Phys. Rev. A {\bf 83},  012503 (2011).

\bibitem{Tchoukova2008}
E. Iskrenova-Tchoukova and M. S. Safronova, Theoretical study of lifetimes and polarizabilities in Ba$^+$, Phys. Rev. A {\bf 78},  012508 (2008).

\bibitem{Jiang2009}
D. Jiang, B. Arora, M. S. Safronova, and C. W. Clark, Blackbody-radiation shift in a $^{88}$Sr$^+$ ion optical frequency standard, J. Phys. B: At. Mol. Opt. Phys. {\bf 42}, 154020 (2009).

\bibitem{Sahoo2009-2}
B. K. Sahoo, B.P. Das, D. Mukherjee, Relativistic coupled-cluster studies of ionization potentials, lifetimes, and polarizabilities in singly ionized calcium, Phys. Rev. A {\bf 79}, 052511 (2009).

\bibitem{Sahoo2009CaSrBa}
B. K. Sahoo, Nuclear quadrupole moment of $^{43}$Ca and hyperfine-structure studies of its singly charged ion, Phys. Rev. A {\bf 80}, 012515 (2009); Comparative studies of dipole polarizabilities in Sr$^+$, Ba$^+$, and Ra$^+$ and their applications to optical clocks, Phys. Rev. A {\bf 73}, 062501 (2009); Relativistic coupled-cluster theory of quadrupole moments and hyperfine structure constants of 5$d$ states in Ba$^+$, Phys. Rev. A {\bf 74}, 020501(R) (2006).
  
\bibitem{Tangarxiv}
Yong-Bo Tang, Precision calculation of hyperfine-structure constants for extracting the nuclear quadrupole moment of $^{43}$Ca, Phys. Rev. A {\bf 112}, 042814 (2025).

\bibitem{Yu2004}
K. Z. Yu, L. J. Wu, B. C. Gou, and T. Y. Shi, Calculation of the hyperfine structure constants in $^{43}$Ca$^+$ and $^{87}$Sr$^+$, Phys. Rev. A {\bf 70}, 012506 (2004).

\bibitem{Pyykko2001}
P. Pyykkö, Spectroscopic nuclear quadrupole moments,  Mol. Phys. {\bf 99}, 1617 (2001).

\bibitem{grasp}
F. A. Parpia and C. Froese Fischer and I. P. Grant, GRASP92: A package for large-scale relativistic atomic structure calculations, Comp. Phys. Comm. {\bf 94}, 249 (1996). 

\bibitem{suppl}
See Supplemental Materials [url] for further discussions and additional results, which include Refs. [52-61].

\bibitem{Manakov1986}
N. L. Manakov, V. D. Ovsiannikov, and L. P. Rapoport, Atoms in a laser field, Phys. Rep. {\bf 141}, 320 (1986).

\bibitem{Stalnaker2006}
J. E. Stalnaker, D. Budker, S. J. Freedman, J. S. Guzman, S. M. Rochester, and V. V. Yashchuk, Dynamic Stark effect and forbidden-transition spectral line shapes, Phys. Rev. A {\bf 73}, 043416 (2006).

\bibitem{Kozlov1999}
M. G. Kozlov and S. G. Porsev, Polarizabilities and hyperfine structure constants of the low-lying levels of barium, Eur. Phys. J. D {\bf 5}, 59 (1999).

\bibitem{Lindgren1986}
I. Lindgren and J. Morrison, {\it Atomic many-body theory}, Springer-Verlag Berlin Heidelberg (1986).

\bibitem{Bartlett2009}
I. Shavitt, R. J. Bartlett, {\it Many-body methods in chemistry and physics: MBPT and coupled-cluster theory}, Cambridge University Press (2009).

\bibitem{Chakraborty2023}
A. Chakraborty and B. K. Sahoo, Deciphering core, valence, and double-core-polarization contributions to parity violating amplitudes in $^{133}$Cs using different many-body methods, J. Phys. Chem. A {\bf 127}, 7518 (2023).

\bibitem{Bohr1950}
A. Bohr and V.F. Weisskopf, The influence of nuclear structure on the hyperfine structure of heavy elements, Phys. Rev. {\bf 77}, 94 (1950).

\bibitem{Shabaev1997}
V. M. Shabaev, M. Tomaselli, T. K\"uhl, A. N. Artemyev, and V. A. Yerokhin, Ground-state hyperfine splitting of high-$Z$ hydrogenlike ions, Phys. Rev. A {\bf 56}, 252 (1997).

\bibitem{Volotka2008}
A. V. Volotka, D. A. Glazov, I. I. Tupitsyn, N. S. Oreshkina, G. Plunien, and V. M. Shabaev, Ground-state hyperfine structure of H-, Li-, and B-like ions in the intermediate-$Z$ region, Phys. Rev. A {\bf 78}, 062507 (2008).

\bibitem{Stone}
N. J. Stone, Table of nuclear magnetic dipole and electric quadrupole moments, At. Data Nucl. Data Tables {\bf 90}, 75 (2005).

\bibitem{Ginges2016}
J. S. M. Ginges, J. C. Berengut, Atomic many-body effects and Lamb shifts in alkali metals, Phys. Rev. A {\bf 93}, 052509 (2016).

\bibitem{Sahoo2016}
B. K. Sahoo, Conforming the measured lifetimes of the 5$d~^2D_{3/2,5/2}$ states in Cs with theory, Phys. Rev. A {\bf 93}, 022503 (2016).

\bibitem{NIST}
A. Kramida, Y. Ralchenko, and J. Reader, National Institute of Standards and Technology, Gaithersburg, MD (Available at: http://physics. nist. gov/asd) (2018).

\bibitem{Chakraborty2025}
A. Chakraborty and B. K. Sahoo, {\it Ab initio} calculations of electric dipole polarizabilities in the Li, Na, and K atoms, Phys. Rev. A {\bf 111}, 042807 (2025).

\bibitem{Chakraborty2025-2}
A. Chakraborty and B. K. Sahoo, Demonstrating correlation trends in the electric dipole polarizabilities of many low-lying states in Cs I through first-principles calculations, Phys. Rev. A {\bf 111}, 062812 (2025).

\bibitem{Safronova2017}
M. S. Safronova, U.I. Safronova, and W. R. Johnson, Forbidden M1 and E2 transitions in monovalent atoms and ions, Phys. Rev. A {\bf 95},  042507 (2017).

\bibitem{Kaur2021}
M. Kaur, D. F. Dar, B. K. Sahoo, and B. Arora, Radiative transition properties of singly charged magnesium, calcium, strontium and barium ions, At. Data Nucl. Data Tables {\bf 137}, 101381 (2021).

\bibitem{Mani2010}
B. K. Mani and D. Angom, Atomic properties calculated by relativistic coupled-cluster theory without truncation: Hyperfine constants of Mg$^+$, Ca$^+$, Sr$^+$, and Ba$^+$, Phys. Rev. A {\bf 81}, 042514 (2010). 

\end{thebibliography}

\begin{thebibliography}{5}

\bibitem{Manakov1986}
N. L. Manakov, V. D. Ovsiannikov, and L. P. Rapoport, Atoms in a laser field, Phys. Rep. {\bf 141}, 320 (1986).

\bibitem{Stalnaker2006}
J. E. Stalnaker, D. Budker, S. J. Freedman, J. S. Guzman, S. M. Rochester, and V. V. Yashchuk, Dynamic Stark effect and forbidden-transition spectral line shapes, Phys. Rev. A {\bf 73}, 043416 (2006).

\bibitem{Kozlov1999}
M. G. Kozlov and S. G. Porsev, Polarizabilities and hyperfine structure constants of the low-lying levels of barium, Eur. Phys. J. D {\bf 5}, 59 (1999).

\bibitem{Lindgren1986}
I. Lindgren and J. Morrison, {\it Atomic many-body theory}, Springer-Verlag Berlin Heidelberg (1986).

\bibitem{Bartlett2009}
I. Shavitt, R. J. Bartlett, {\it Many-body methods in chemistry and physics: MBPT and coupled-cluster theory}, Cambridge University Press (2009).

\bibitem{Chakraborty2023}
A. Chakraborty and B. K. Sahoo, Deciphering core, valence, and double-core-polarization contributions to parity violating amplitudes in $^{133}$Cs using different many-body methods, J. Phys. Chem. A {\bf 127}, 7518 (2023).

\bibitem{Bohr1950}
A. Bohr and V.F. Weisskopf, The influence of nuclear structure on the hyperfine structure of heavy elements, Phys. Rev. {\bf 77}, 94 (1950).

\bibitem{Shabaev1997}
V. M. Shabaev, M. Tomaselli, T. K\"uhl, A. N. Artemyev, and V. A. Yerokhin, Ground-state hyperfine splitting of high-$Z$ hydrogenlike ions, Phys. Rev. A {\bf 56}, 252 (1997).

\bibitem{Volotka2008}
A. V. Volotka, D. A. Glazov, I. I. Tupitsyn, N. S. Oreshkina, G. Plunien, and V. M. Shabaev, Ground-state hyperfine structure of H-, Li-, and B-like ions in the intermediate-$Z$ region, Phys. Rev. A {\bf 78}, 062507 (2008).

\bibitem{Stone}
N. J. Stone, Table of nuclear magnetic dipole and electric quadrupole moments, At. Data Nucl. Data Tables {\bf 90}, 75 (2005).


\end{thebibliography}
\end{document}